%% file: biclustering_20210608.tex
\documentclass[12pt]{article}

\usepackage{times,epsfig,graphics,amssymb,latexsym,amsmath,setspace}%,fullpage}
\usepackage{array,threeparttable,hyperref,url,longtable}
\usepackage[usenames]{color}
\usepackage[authoryear,sort]{natbib}
\usepackage{paralist}
\usepackage{mathabx}
\usepackage{algorithm,algorithmic}
\usepackage{ntheorem}
\usepackage[normalem]{ulem}
\usepackage{authblk}
\usepackage[symbol*]{footmisc}
\usepackage{tabularx,ragged2e}
\usepackage{xr}
\usepackage{booktabs}
\usepackage{xcolor}
\usepackage{xparse}
\usepackage{listings}
\usepackage{chngcntr}

\counterwithout{table}{section}

\allowdisplaybreaks

\input self-define-YF.tex

\def\boxit#1{\vbox{\hrule\hbox{\vrule\kern6pt\vbox{\kern6pt#1\kern6pt}\kern6pt\vrule}\hrule}}

\catcode`@=11
\def\@evenhead{\vbox{\hbox to \textwidth{\tiny \hfill \hfill \today } }}
\def\@oddhead{\vbox{\hbox to \textwidth{\tiny \hfill \hfill \today } }}

\def\argmin{\mathop{\rm argmin}}

\def\author.arg{
\medskip
Binhuan Wang \\
Department of Population Health\\
New York University School of Medicine\\

\medskip
Lanqiu Yao \\
Department of Population Health\\
New York University School of Medicine\\

\medskip
Jiyuan Hu \\
Department of Population Health\\
New York University School of Medicine\\

\medskip
Huilin Li \\
Department of Population Health\\
New York University School of Medicine\\
}

\def\tit.arg{A New Algorithm for Convex Biclustering and \\ Its Extension to the Compositional Data}

\begin{document}
\pagenumbering{arabic}
\setcounter{page}{1}
\baselineskip=14pt

\begin{center}
{\Large \tit.arg} \\

\vskip 3mm

\author.arg
\end{center}

\vskip 3mm

\date{}

\begin{abstract}
Biclustering is a powerful data mining technique that allows simultaneously clustering rows (observations) and columns (features) in a matrix-format data set, which can provide results in a checkerboard-like
pattern for visualization and exploratory analysis in a wide array of domains. Multiple biclustering algorithms have been developed in the past two decades, among which the convex biclustering can guarantee a global optimum by formulating in as a convex optimization problem. On the other hand, the application of biclustering has not progressed in parallel with the algorithm techniques. For example, biclustering for increasingly popular microbiome research data is under-applied possibly due to its compositional constraints for each sample. In this manuscript, we propose a new convex biclustering algorithm, called the bi-ADMM, under general setups based on the ADMM algorithm, which is free of extra smoothing steps to visualize informative biclusters required by existing convex biclustering algorithms. Furthermore, we tailor it to the algorithm named biC-ADMM specifically to tackle compositional constraints confronted in microbiome data. The key step of our methods utilizes the {\it Sylvester Equation} to derive the ADMM algorithm, which is new to the clustering research. The effectiveness of the proposed methods is examined through a variety of numerical experiments and a microbiome data application.

\end{abstract}


\doublespacing

\section{Introduction}

Biclustering is a powerful data mining technique that allows simultaneously clustering rows (observations) and columns (features) in a matrix-format data set. It is a 2-way extension of clustering analysis which aims to assign observations into a number of clusters such that observations in the same group are similar to each other. Biclustering can provide results in a checkerboard-like pattern, which can be used for visualization and exploratory analysis in a wide array of domains. For example, \cite{cheng2000biclustering} introduced the concept of biclustering for the first time to gene expression data, aiming to overcome some problems of traditional clustering
methods such as information loss during oversimplied similarity and grouping computation. Instead, biclustering identified genes expression patterns under a subset of all the conditions/samples, which provided a better representation for genes with multiple functions or regulated by many factors. Furthermore, in the big data era, electronic health records contain abundant information that can be transformed into disease phenotypes. These phenotype data can be organized into a matrix, with individuals as rows and phenotype features as columns, from which biclustering can discover a subgroup of patients from a subset of phenotypes \citep{hripcsak2012next}. Results from biclustering can shed lights on the downstream analyses to interpret the identified biclusters and to demonstrate bicluster associations coupled with other statistical evaluations tool.

In the past two decades, multiple biclustering algorithms have been developed. The most popular approach for genomic data is the clustered dendrogram \citep{madeira2004biclustering,busygin2008biclustering}, which actually performs the hierarchical clustering \citep{hastie2009} on both subjects and genes. However, dendrogram is less ideal for generating reproducible results because it greedily fuses observations or features to minimize a specific objective function such as the within-clustering variance. As a consequence, this algorithm may return a locally optimal solution with respect to the criterion, and dendrogram is not stable to small perturbations of the data due to its greedy algorithm. Later on, more advanced biclustering methods have been proposed, which are either based on the singular value decomposition (SVD) or graph cuts. However, \cite{chi2017biconvex} pointed out that none of these methods addressed the fundamental issues that underlay the clustered dendrogram. Alternatively, they formulated the biclustering problem as a convex optimization problem, and solved it with an iterative algorithm based on the splitting methods for convex clustering.

Existing biclustering methods have several ways to define biclusters: 1) defining a submatrix with a large average as a bicluster \citep{shabalin2009finding}; 2) defining a bicluster by coherent low-rank patterns \citep{Lee2010,li2020generalized}; 3) assuming the elements of a bicluster which shares the same underlying mean \citep{flynn2012consistent,Tan2014,chi2017biconvex}. To be specific, this manuscript adopts the third way as it is motivated by \cite{chi2017biconvex,Wang2018sparse}.

On the other hand, application of biclustering has not progressed in parallel with the algorithmic development. Recently, \cite{xie2018time} provided a comprehensive review of biclustering applications in biological and biomedical data, and pointed out there is a need for developing supporting computational techniques and providing the guidance on selecting the appropriate biclustering tools in a specific study. Data generated from different biotechnologies have their own properties. For example, 16S and shotgun sequencing are two widely used sequencing techniques in microbiome studies. The raw microbiome data are typically organized into large matrices with rows representing samples, and columns containing observed counts of clustered sequences commonly known as operational taxonomic units (OTUs).  Because the observed counts are not comparable across samples due to their constrained total by the sequencing depth \citep{gloor2017microbiome}, normalizing counts toward their totals into the relative abundance is a commonly used metric in microbiome studies, which imposes a compositional constraint for data from each sample. As reviewed by \cite{xie2018time}, there are limited applications of biclustering in microbiome studies, except one example. \cite{falony2016population} identified sample subsets with specific taxonomic signatures using a biclustering approach called FABIA, proposed by \cite{hochreiter2010fabia}.

In this manuscript, we propose a new convex biclustering algorithm, named bi-ADMM, under general setups based on the alternating direction method of multipliers (ADMM) algorithm. Furthermore, we tailor it to the algorithm, named biC-ADMM, specifically to tackle the compositional constraints confronted in microbiome data. As far as we have known, it is the first biclustering algorithm that can handle data with compositional constraints. Compared with \cite{chi2017biconvex}, the novelty of this paper is four-fold. First, we solve the convex biclustering problem via the ADMM algorithm directly. As \cite{chi2017biconvex} mentioned, although an ADMM algorithm was feasible, they solved it in an alternative way as a convex bicluster ring algorithm (COBRA) via iteratively using Dykstra-like proximal algorithm (DLPA) combined with the alternating minimization algorithm (AMA) introduced in \cite{Chi2015} as the convex clustering solver. In the meanwhile, their methods require extra smoothing steps to facilitate the visualization of estimtaed biclusters, seeing Figure 1(b) in \cite{chi2017biconvex}. The key step of our methods utilizes the {\it Sylvester Equation} for the first time to solve the challenging estimation problem involving constraints on both rows and columns of a matrix. Second, the proposed algorithms can naturally incorporate additional constraints for rows or columns. For example, the modified algorithms for the microbiome data with compositional constraint for data rows are provided. Third, the proposed algorithms use two tuning parameters to control the resulting numbers of row and column clusters separately, which includes the single parameter setting used by \cite{chi2017biconvex} to controls the biclustering path as a special case. As a result, our methods offer users an opportunity to fix clusters for one dimenstion and monitor the clustering path of the other one. Fourth, the fused-lasso penalty terms in the proposed methods accommodate $L_q$-norm, $q=1,2,\infty$, and thus, are more flexible than the COBRA which requires the $L_2$-norm. Note that our methods are not only theoretically sound, but also practically promising. The superior performance of our methods is demonstrated in extensive simulated examples and the application to analyze a murine gut microbiome dataset \citep{livanos2016antibiotic}.

\begin{figure}[!htb]
\protect\caption{The heat maps of estimated matrices based on the COBRA (left) and the bi-ADMM (right) algorithms with the same tuning parameter $\gamma=6.72$, respectively. Rows indicate 80 samples with 4 clusters and columns indicate 40 features with 4 clusters. The row(column) side color bars indicate the true row(column) clusters.}
\centering \includegraphics[scale=0.5]{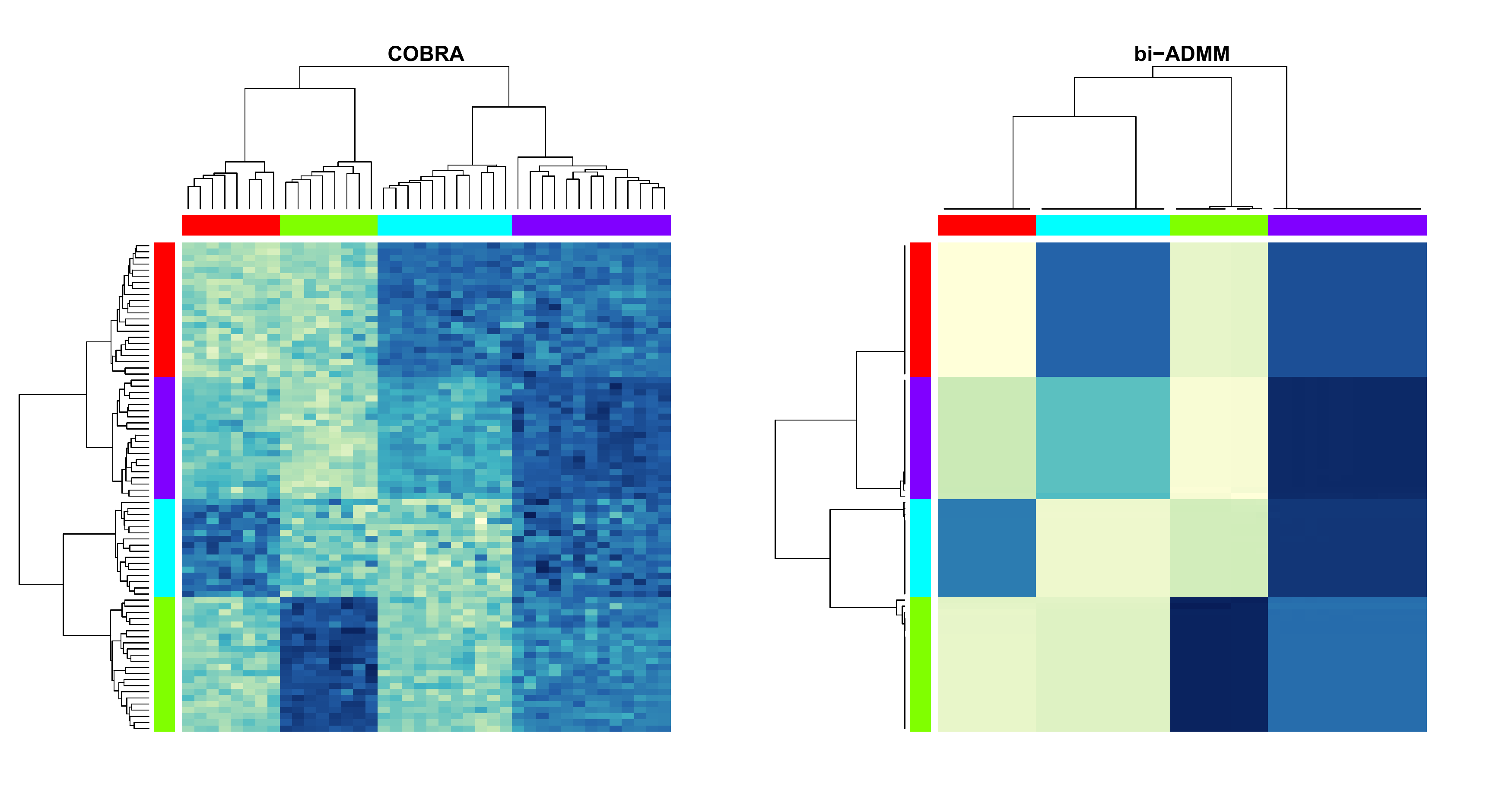}\label{comparison}
\end{figure}

We demonstrate the superior performance of the proposed algorithm under the simulation setting in Section \ref{sec:sim_general} with $\sigma=4$, where $\sigma$ is the standard deviation of the random noise for data generation. To ensure a fair comparison, we adopt a special case of our methods using the objective function \eqref{obj_general_single} with a single tuning parameter, which is also used by COBRA. In this setting, there are $n=80$ samples from 4 clusters and $p=40$ features from 4 clusters. Figure \ref{comparison} compares the performance of COBRA without the smoothing step (indicated as COBRA on the left panel) and our method (indicated as bi-ADMM on the right panel), by visualizing the estimate for each method with the same set of weights and the same value for the single tuning parameter, $\gamma=6.72$. The heat maps show that the proposed bi-ADMM algorithm can directly visualize the estimate with a clear checkerboard-like pattern without any further procedure.

The rest of the paper is organized as follows. We propose the main methods and algorithms in Section 2. Some practical consideraions for applying the proposed algorithms are discussed in Section 3. The performance of the proposed methods via simulation studies are evaluated in Section 4, and a microbiome application is presented in Section 5. We conclude the paper with a brief summary in Section 6, and defer all technical proofs to the Appendix.

\section{Convex Biclustering}

In this section, we develop algorithms for a general biclustering problem first, and then extend them to incorporate the compositional constraints, a unique feature of the microbiome data.

\subsection{Specification of the General Model} \label{sec:genera_model}

We consider a general biclustering problem as follows. Let $\mbfX\in\mathbb{R}^{n\times p}$ be a data matrix with $n$ observations $X_{i\cdot}=(X_{i1},\ldots,X_{ip})\trans$ with $p$ features, $i=1,\cdots,n$. The underlying assumption for applying a biclustering method is that the $n$ observations belong to $K$ unknown and non-overlapping classes, $C_1,\ldots, C_K$, and the $p$ features belong to $R$ unknown and non-overlapping classes, $D_1,\ldots, D_R$. Such checkerboard-like structure of a data matrix motivates researchers to use the fused-lasso penalty \citep{Tibshirani.ea:2005} to fuse rows and columns simultaneously.

To facilitate further derivations, we rewrite the data matrix $\mbfX$ in feature-level as column vectors $\mbfx_j$ so that $\mbfX=(\mbfx_{1},\cdots,\mbfx_{p})$, where $\mbfx_{j}=(X_{1j},\cdots,X_{nj}){\trans}$, $j=1,\ldots,p$. Similarly we denote $\mbfA$ in the feature-level consisting of column vectors $\mathbf{A}=(\mathbf{a}_{1},\cdots,\mathbf{a}_{p})$, where $\mbfa_{j}=(A_{1j},\cdots,A_{nj}){\trans}$, and as $(A_{1\cdot},\ldots, A_{n\cdot})\trans$ in the observation-level. Define $\calE_1=\{l=(l_1,l_2):1\leq l_1<l_2 \leq n\}$ and $\calE_2=\{k=(k_1,k_2):0\leq k_1<k_2 \leq p \}$. Then denote $|\calE_1|$ and $|\calE_2|$ as the numbers of components of $\calE_1$ and $\calE_2$, respectively.

We formulate the convex biclustering problem as the following minimization problem with tuning parameters $\gamma_1$ and $\gamma_2$:
\begin{eqnarray} \label{obj_general}
\min_{\mbfA \in \mathbb{R}^{p \times n}} && \frac{1}{2} \sum_{i=1}^n\| X_{i\cdot}- A_{i\cdot} \|_2^2  +
\gamma_1 \sum_{l \in \calE_1} w_l \| A_{l_1\cdot} -A_{l_2\cdot} \|_q + \gamma_2 \sum_{k \in \calE_2} u_k \| \mbfa_{k_1} -\mbfa_{k_2} \|_q,
\end{eqnarray}
where the weights $w_{l}\ge0$ and $u_{k}\ge0$, and $\| \cdot \|_q$ is the $L_q$-norm of a vector with $q\in \{1,2,\infty\}$. \cite{chi2017biconvex} and \cite{Wang2018sparse} suggested $w_{l}=\iota_{l_{1},l_{2}}^{m_1}\exp(-\phi\|X_{l_{1}\cdot}-X_{l_{2}\cdot}\|_{2}^{2})$,
where $\iota_{l_{1},l_{2}}^{m_1}$ is 1 if observation $l_{2}$ is among
$l_{1}$'s $m_1$ nearest neighbors or vice verse, and 0 otherwise. Similarly, we suggest using $u_{k}=\iota_{k_{1},k_{2}}^{m_2}\exp(-\phi\|\mbfx_{k_{1}}-\mbfx_{k_{2}}\|_{2}^{2})$.

By incorporating the fused-lasso penalty in the second and third terms of (\ref{obj_general}),
the above formulation encourages that some of the rows or columns of the solution $\widehat{\bfA}$
are identical. If $\widehat{A}_{l_{1}\cdot}=\widehat{A}_{l_{2}\cdot}$,
then observation $l_{1}$ and observation $l_{2}$ will be assigned
to the same observation (row) cluster; if $\widehat\mbfa_{k_1}=\widehat\mbfa_{k_2}$,
then feature $k_{1}$ and feature $k_{2}$ will be assigned
to the same feature (column) cluster.  The tuning parameter $\gamma_1$ in (\ref{obj_general})
controls the number of unique rows of $\widehat{\bfA}$, that is,
the number of estimated observation clusters, while the tuning parameter $\gamma_2$
controls the number of estimated feature clusters. When $\gamma_1=\gamma_2=0$, $\widehat{\bfA}={\bfX}$,
and thus, each combination of feature and observation by itself is a bicluster. As $\gamma_1$ and $\gamma_2$
increases, some of the rows and columns of $\widehat{\bfA}$ become identical, respectively,
which demonstrates a fusing process. For sufficiently large $\gamma_1$ and $\gamma_2$,
all rows or columns of $\widehat{\bfA}$ will be identical, implying that
all observations or features are estimated to form a single bicluster.
With such a strictly convex objective function in (\ref{obj_general}), the solution $\widehat{\bfA}$ from convex biclustering is unique for each given pair of $\gamma_1$ and $\gamma_2$.

\subsection{bi-ADMM: a Biclustering Algorithm for the General Model} \label{Alg_general}

In order to implement the ADMM algorithm,  we rewrite the above problem into the following constrained optimization problem:
\begin{eqnarray} \label{obj_min}
\min_{\mbfA \in \mathbb{R}^{p \times n}} && \frac{1}{2} \sum_{i=1}^n\| X_{i\cdot}- A_{i\cdot} \|_2^2 + \gamma_1 \sum_{l \in \calE_1} w_l \| \mbfv_l \|_q + \gamma_2 \sum_{k \in \calE_2} u_k \| \mbfz_k \|_q \\
 {\rm s.t.} && A_{l_1\cdot} - A_{l_2\cdot} -\mbfv_l = \bfzero, \ \forall l \in \calE_1 \nonumber \\
 &&  \mbfa_{k_1} - \mbfa_{k_2} -\mbfz_k = \bfzero, \ \forall k \in \calE_2, \nonumber
\end{eqnarray}
which is also equivalent to an augmented Lagrangian problem given by
\begin{eqnarray} \label{obj_augL}
\mathcal{L}_{\nu_1,\nu_2}(\mbfA,\mbfV,\mbfZ,\bLambda_1,\bLambda_2) &=&  \frac{1}{2} \sum_{i=1}^n\| X_{i\cdot}- A_{i\cdot} \|_2^2
+ \gamma_1 \sum_{l \in \calE_1} w_l \| \mbfv_l \|_q  + \gamma_2 \sum_{k \in \calE_2} u_k \| \mbfz_k \|_q \nonumber \\
&& + \sum_{l \in \calE_1} \langle\blambda_{1l}, \mbfv_l -A_{l_1\cdot}+A_{l_2\cdot}\rangle+ \frac{\nu_1}{2} \sum_{l \in \calE_1} \| \mbfv_l -A_{l_1\cdot}+A_{l_2\cdot} \|_2^2  \nonumber\\
&&  + \sum_{k \in \calE_2} \langle\blambda_{2k}, \mbfz_k -\mbfa_{k_1}+\mbfa_{k_2}\rangle + \frac{\nu_2}{2} \sum_{k \in \calE_2}\| \mbfz_k -\mbfa_{k_1}+\mbfa_{k_2} \|_2^2,
\end{eqnarray}
where $\nu_1$ and $\nu_2$ are nonnegative constants, $\mbfV=(\mbfv_{1},\ldots,\mbfv_{|{\calE}_1|})$, $\mbfZ=(\mbfz_{1},\ldots,\mbfz_{|{\calE}_2|})$, $\bLambda_1=(\blambda_{11},\ldots,\blambda_{1|{\calE}_1|})$ and $\bLambda_2=(\blambda_{21},\ldots,\blambda_{2|{\calE}_2|})$. This optimization problem is challenging in dealing with fusing both observation-level and feature-level vectors in the objective function simultaneously.

The bi-ADMM minimizes the augmented Lagrangian problem by alternatively
solving one block of variables at a time. Specifically, bi-ADMM solves
\begin{eqnarray}
\mbfA^{m+1} & = & \argmin_{\mbfA}\calL_{\nu_1,\nu_2}(\mbfA,\mbfV^m,\mbfZ^m,\bLambda_1^m,\bLambda_2^m),\nonumber \\
\mbfV^{m+1} & = & \argmin_{\mbfV}\calL_{\nu_1,\nu_2}(\mbfA^{m+1},\mbfV,\mbfZ^m,\bLambda_1^m,\bLambda_2^m), \nonumber\\
\mbfZ^{m+1} & = & \argmin_{\mbfZ}\calL_{\nu_1,\nu_2}(\mbfA^{m+1},\mbfV^{m+1},\mbfZ,\bLambda_1^m,\bLambda_2^m),\\
\blambda_{1l}^{m+1} & = & \blambda_{1l}^{m}+\nu_1(\mbfv_{l}^{m+1}- A_{l_{1}\cdot}^{m+1}+A_{l_{2}\cdot}^{m+1}), \ l \in \calE_1,\nonumber \\
\blambda_{2k}^{m+1} & = & \blambda_{2k}^{m}+\nu_2(\mbfz_{k}^{m+1}- \mbfa_{k_{1}}^{m+1}+\mbfa_{k_{2}}^{m+1}), \ k \in \calE_2.\nonumber
\end{eqnarray}

Next, we develop the detailed updating implementations for $\mbfA,\mbfV,\mbfZ,\bLambda_1$
and $\bLambda_2$ in three steps. A summary of the bi-ADMM algorithm is shown in Algorithm \ref{algbiADMM}.

\textbf{Step 1: update $\mbfA$.} To update $\mbfA$, we need to minimize
\begin{eqnarray} \label{eqn:bi_update_A}
f(\mbfA)&=&\frac{1}{2} \sum_{i=1}^n\| X_{i\cdot}- A_{i\cdot} \|_2^2 + \frac{\nu_1}{2} \sum_{l \in \calE_1} \| \wt{\mbfv}_l-A_{l_1 \cdot} + A_{l_2 \cdot} \|_2^2
 +\frac{\nu_2}{2} \sum_{k \in \calE_2} \| \wt{\mbfz}_k -\mbfa_{k_1}+\mbfa_{k_2} \|_2^2,
\end{eqnarray}
where $\wt{\mbfv}_1=\mbfv_l + \frac{1}{\nu_1}\blambda_{1l}$ and $\wt{\mbfz}_k=\mbfz_k + \frac{1}{\nu_2}\blambda_{2k}$.

The optimization problem is challenging because it involves both rows and columns of a single matrix. To tackle this difficulty, the following key lemma associating $(\ref{eqn:bi_update_A})$ with a {\it Sylvester Equations} is proposed. The proof of Lemma \ref{thm:biADMM} is provided in Appendix.

\begin{lemma} \label{thm:biADMM}
Let $\bfI_{n}$ be an $n\times n$ identity matrix, $\bfone_{n}$ be an $n$-dimensional
vector with each component being 1, $\mbfe_{i}$ be an $n$-dimensional
vector with all components being 0, except the $i$-th component being
1, and $\mbfe_{j}^*$ is a $p$-dimensional vector with the $j$-th element as 1 and 0 otherwise. In addition, define
\begin{eqnarray*}
\mbfM &=&  \mbfI_n + \nu_1 \sum_{l \in \calE_1} (\mbfe_{l_1}-\mbfe_{l_2})(\mbfe_{l_1}-\mbfe_{l_2})\trans \\
\mbfN &=& \nu_2 \sum_{k \in \calE_2} (\mbfe_{k_1}^*-\mbfe_{k_2}^*) (\mbfe_{k_1}^*-\mbfe_{k_2}^*)\trans \\
\mbfG&=& \mbfX + \sum_{l \in \calE_1} (\mbfe_{l_1}-\mbfe_{l_2})(\blambda_{1l}+ \nu_1\mbfv_l)\trans + \sum_{k \in \calE_2} (\blambda_{2k}+\nu_2\mbfz_k)(\mbfe_{k_1}^*-\mbfe_{k_2}^*)\trans.
\end{eqnarray*}
Then,
minimizing $(\ref{eqn:bi_update_A})$ is equivalent to solving the following {\it Sylvester Equation}
\begin{eqnarray}\label{bi_Stlvesterequation}
\mbfM\mbfA + \mbfA\mbfN = \mbfG.
\end{eqnarray}
\end{lemma}

If the edge sets $\calE_1$ and $\calE_2$ contain all possible edges, it is straightforward to verify
\begin{eqnarray*}
\sum_{l \in \calE_1} (\mbfe_{l_1}-\mbfe_{l_2})(\mbfe_{l_1}-\mbfe_{l_2})\trans &=& n\mbfI_n - \bfone_n \bfone_n\trans \\
\sum_{k \in \calE_2} (\mbfe_{k_1}^*-\mbfe_{k_2}^*) (\mbfe_{k_1}^*-\mbfe_{k_2}^*)\trans &=& p\mbfI_p - \bfone_p \bfone_p\trans.
\end{eqnarray*}
Then, $\mbfM = (1+n\nu_1)\mbfI_n -\nu_1 \bfone_n \bfone_n\trans$ and $\mbfN = p\nu_2 \mbfI_p - \nu_2 \bfone_p \bfone_p\trans$.

The type of equation \eqref{bi_Stlvesterequation} is called {\it Sylvester Equation}, which is very important
in control theory and many other branches of engineering. Its theoretical solution is based on eigenvector and eigenvalue decomposition \citep{Jameson1968}, but it is computationally expensive. The Bartels-Stewart algorithm \citep{bartels1972solution} is the standard numerical solution that transforms the {\it Sylvester Equation} into a triangular system with the Schur decomposition and then solves it with forward or backward substitutions. In this manuscript and its accompanied $\textsf{R}$ package, we implement a modified Bartels-Stewart algorithm proposed by \cite{Sorensen2003}, which is more efficient.

\textbf{Step 2: Update $\mbfV$ and $\mbfZ$.} For any $\sigma>0$ and norm $\Omega(\cdot)$,
we define a proximal map,
\begin{eqnarray*}
\textrm{prox}_{\sigma\Omega}(\bfu)=\argmin_{\bfv}\left[\sigma\Omega(\bfv)+\frac{1}{2}\|\bfu-\bfv\|_{2}^{2}\right].
\end{eqnarray*}
We refer the readers to Table 1 in \citet{Chi2015} for the solutions to
the proximal map of $L_q$-norm for $q=1,2$ and $\infty$. It is clear that the vectors ${\mbfv}_l$ and ${\mbfz}_k$ are separable in the objective function \eqref{obj_augL}, and thus ${\mbfv}_l$ and ${\mbfz}_k$ can be solved via proximal maps:
\begin{eqnarray*}
\mbfv_l &=& \argmin_{\mbfv_l} \frac{1}{2} \|\mbfv_l - (A_{l_1\cdot}-A_{l_2\cdot} - \nu_1^{-1}\blambda_{1l}) \|_2^2
 + \frac{\gamma_1 w_l}{\nu_1} \| \mbfv_l\|_q \\
 &=& \textrm{prox}_{\sigma_{1l}\|\cdot\|_q} (A_{l_1\cdot}-A_{l_2\cdot}- \nu_1^{-1}\blambda_{1l}) \\
\mbfz_k &=& \argmin_{\mbfz_k} \frac{1}{2} \|\mbfz_k - (\mbfa_{k_1}-\mbfa_{k_2} - \nu_2^{-1}\blambda_{2k}) \|_2^2
 + \frac{\gamma_2 u_k}{\nu_2} \| \mbfz_k\|_q \\
 &=& \textrm{prox}_{\sigma_{2k}\|\cdot\|_q} (\mbfa_{k_1}-\mbfa_{k_2} - \nu_2^{-1}\blambda_{2k}),
\end{eqnarray*}
where $\sigma_{1l}=\gamma_1 w_l/\nu_1$ and $\sigma_{2k}=\gamma_2 u_k/\nu_2$.

\textbf{Step 3: Update $\bLambda_1$ and $\bLambda_2$.} Finally, $\blambda_{1l}$ and $\blambda_{2k}$ can be updated by $\blambda_{1l} = \blambda_{1l} + \nu_1(\mbfv_l - A_{l_1\cdot} + A_{l_2\cdot})$ and $\blambda_{2k} = \blambda_{2k} + \nu_2(\mbfz_k - \mbfa_{k_1}+\mbfa_{k_2})$.

\begin{algorithm}[!htb]
\protect\caption{\quad{}bi-ADMM \label{algbiADMM}}

\begin{enumerate}
\item Initialize $\mbfV^{0}, \mbfZ^{0}, \bLambda_1^{0}$ and $\bLambda_2^{0}$. Calculate
\begin{eqnarray*}
\mbfM &=&  \mbfI_n + \nu_1 \sum_{l \in \calE_1} (\mbfe_{l_1}-\mbfe_{l_2})(\mbfe_{l_1}-\mbfe_{l_2})\trans \\
\mbfN &=& \nu_2 \sum_{k \in \calE_2} (\mbfe_{k_1}^*-\mbfe_{k_2}^*) (\mbfe_{k_1}^*-\mbfe_{k_2}^*)\trans
\end{eqnarray*}
For $m=1,2,\ldots$
\item Solve the {\it Sylvester Equation} $\mbfM\mbfA + \mbfA\mbfN = \mbfG^{m-1}$ to obtain $\mbfA^m$, where
\begin{eqnarray*}
\mbfG^{m-1}= \mbfX + \sum_{l \in \calE_1} (\mbfe_{l_1}-\mbfe_{l_2})(\blambda_{1l}^{m-1}+ \nu_1\mbfv_l^{m-1})\trans + \sum_{k \in \calE_2} (\blambda_{2k}^{m-1}+\nu_2\mbfz_k^{m-1})(\mbfe_{k_1}^*-\mbfe_{k_2}^*)\trans.
\end{eqnarray*}
\item For $l\in\calE_1$, do
\begin{eqnarray*}
\mbfv_{l}^{m}= \textrm{prox}_{\sigma_{1l}\|\cdot\|_q} (A_{l_1\cdot}^{m}-A_{l_2\cdot}^{m}- \nu_1^{-1}\blambda_{1l}^{m-1}).
\end{eqnarray*}
\item For $k\in\calE_2$, do
\begin{eqnarray*}
\mbfz_{l}^{m}= \textrm{prox}_{\sigma_{2k}\|\cdot\|_q} (\mbfa_{k_1}^m-\mbfa_{k_2}^m - \nu_2^{-1}\blambda_{2k}^{m-1}).
\end{eqnarray*}
\item For $l\in\calE_1$ and $k \in\calE_2$, do
\begin{eqnarray*}
\blambda_{1l}^m &=& \blambda_{1l}^{m-1} + \nu_1(\mbfv_l^m - A_{l_1\cdot}^m + A_{l_2\cdot}^m)\\
\blambda_{2k}^m &=& \blambda_{2k}^{m-1} + \nu_2(\mbfz_k^m - \mbfa_{k_1}^m+\mbfa_{k_2}^m)
\end{eqnarray*}

\item Repeat Steps 2-5 until convergence. \end{enumerate}
\end{algorithm}

\subsection{biC-ADMM: Biclustering Algorithm for Data with Compositional Constraints} \label{Alg_compositional}

For microbiome data, due to its data-generating mechanism, the counts of all taxa need to be normalized into the relative abundances such that their summation for each sample is 1. This type of data is also called compositional data in mathematics. Our newly proposed clustering algorithm, biC-ADMM, aims to incorporate the compositional constraints into the optimization. To be specific, the data from each sample satisfy $X_{i\cdot}\trans \bfone_p = 1, i=1,\ldots,n$, and then it is natural to require the estimate $\mbfA$ to satisfy the same constraints, i.e., $A_{i\cdot}\trans \bfone_p = 1$.

Because the bi-ADMM solves an optimization problem with constraints, it can be easily extended to the biC-ADMM algorithm by incorporating the compositional constraints. With the additional compositional constraints into \eqref{obj_min}, the augmented Lagrangian problem for compositional data is given by
\begin{eqnarray} \label{obj_augL_c}
\mathcal{L}_{\nu_1,\nu_2,\nu_3}(\mbfA,\mbfV,\mbfZ,\bLambda_1,\bLambda_2,\blambda_3) &=&  \frac{1}{2} \sum_{i=1}^n\| X_{i\cdot}- A_{i\cdot} \|_2^2
+ \gamma_1 \sum_{l \in \calE_1} w_l \| \mbfv_l \|_q  + \gamma_2 \sum_{k \in \calE_2} u_k \| \mbfz_k \|_q \nonumber \\
&& + \sum_{l \in \calE_1} \langle\blambda_{1l}, \mbfv_l -A_{l_1\cdot}+A_{l_2\cdot}\rangle+ \frac{\nu_1}{2} \sum_{l \in \calE_1} \| \mbfv_l -A_{l_1\cdot}+A_{l_2\cdot} \|_2^2  \nonumber\\
&&  + \sum_{k \in \calE_2} \langle\blambda_{2k}, \mbfz_k -\mbfa_{k_1}+\mbfa_{k_2}\rangle + \frac{\nu_2}{2} \sum_{k \in \calE_2}\| \mbfz_k -\mbfa_{k_1}+\mbfa_{k_2} \|_2^2 \nonumber \\
&&  + \langle \blambda_{3},\bfone_n - \mbfA\bfone_p\rangle + \frac{\nu_3}{2} \| \bfone_n - \mbfA\bfone_p \|_2^2,
\end{eqnarray}
where $\nu_3$ is a nonnegative constant and $\blambda_3 = (\lambda_{31},\ldots,\lambda_{3n})\trans$.

As a variant of the bi-ADMM for biclustering compositional data, the biC-ADMM algorithm is similar with the bi-ADMM detailed in Section \ref{Alg_general}. The key step for updating $\mbfA$ of the biC-ADMM is summarized in the following Lemma \ref{thm:biCADMM}, which minimizes:
\begin{eqnarray}\label{eqn:biC_update_A}
f(\mbfa)=\frac{1}{2} \| \mbfx- \mbfa \|_2^2 + \frac{\nu_1}{2} \| \mbfB \mbfP \mbfa - \wt{\mbfv}  \|_2^2  +\frac{\nu_2}{2}  \| \mbfC \mbfa -\wt{\mbfz} \|_2^2 + \frac{\nu_3}{2}  \| \mbfD \mbfa -\bfs \|_2^2,
\end{eqnarray}
with $\mbfx=\textrm{vec}(\mbfX)$, $\mbfa=\textrm{vec}(\mbfA)$, the vectorization of the matrices $\mbfX$ and $\mbfA$, and
\begin{eqnarray*}
&&\mbfB\trans=\left(\mbfB\trans_1,\ldots,\mbfB\trans_{|\calE_1|} \right), \quad \wt{\mbfv}\trans= \left(\wt{\mbfv}_1\trans,\ldots,\wt{\mbfv}_{|\calE_1|}\trans \right), \\
&&\mbfC\trans=\left(\mbfC\trans_1,\ldots,\mbfC\trans_{|\calE_2|} \right), \quad \wt{\mbfz}\trans=
\left( \wt{\mbfz}_1\trans,\ldots,\wt{\mbfz}_{|\calE_2|}\trans \right), \\
&&\mbfD\trans=\left(\mbfD\trans_1,\ldots,\mbfD\trans_n \right), \quad \bfs =\left( s_1,\ldots,s_n \right)\trans=\bfone_n + \blambda_{3}/\nu_3,
\end{eqnarray*}
where $\mbfB_l=(\mbfe_{l_1}-\mbfe_{l2})\trans \otimes \mbfI_p, \mbfC_k=(\mbfe_{k_1}^*-\mbfe_{k2}^*)\trans \otimes \mbfI_n, \mbfD_i= \bfone_p\trans \otimes \mbfe_i\trans$, and $\mbfP$ is a permutation matrix such that $\textrm{vec}(\mbfA\trans)=\mbfP \textrm{vec}(\mbfA), l=1,\ldots,|\calE_1|, k=1,\ldots, |\calE_2|, i=1,\ldots, n$, respectively.

\begin{lemma} \label{thm:biCADMM}
Define
\begin{eqnarray*}
\mbfM &=&  \mbfI_n + \nu_1 \sum_{l \in \calE_1} (\mbfe_{l_1}-\mbfe_{l_2})(\mbfe_{l_1}-\mbfe_{l_2})\trans \\
\mbfN &=& \nu_2 \sum_{k \in \calE_2} (\mbfe_{k_1}^*-\mbfe_{k_2}^*) (\mbfe_{k_1}^*-\mbfe_{k_2}^*)\trans + \nu_3 \bfone_p \bfone_p\trans \\
\mbfG&=& \mbfX + \sum_{l \in \calE_1} (\mbfe_{l_1}-\mbfe_{l_2})(\blambda_{1l}+ \nu_1\mbfv_l)\trans + \sum_{k \in \calE_2} (\blambda_{2k}+\nu_2\mbfz_k)(\mbfe_{k_1}^*-\mbfe_{k_2}^*)\trans + \nu_3 \mbfs \bfone_p\trans.
\end{eqnarray*}
Then,
minimizing $(\ref{eqn:biC_update_A})$ is equivalent to solving the following {\it Sylvester Equation}
\begin{eqnarray}\label{biC_Stlvesterequation}
\mbfM\mbfA + \mbfA\mbfN = \mbfG.
\end{eqnarray}
\end{lemma}

Proof of Lemma \ref{thm:biCADMM} is provided in the Appendix, and the biC-ADMM algorithm is summarized in Algorithm \ref{algbiCADMM}. %and \ref{algbiCAMA}.
%Similarly, setting $\nu_1=\nu_2=\nu_3=0$ for updating $\mbfA$ leads to the biC-AMA algorithm.

\begin{algorithm}[!htb] \scriptsize
\protect\caption{\quad{}biC-ADMM \label{algbiCADMM}}

\begin{enumerate}
\item Initialize $\mbfV^{0}, \mbfZ^{0}, \bLambda_1^{0},\bLambda_2^{0}$ and $\blambda_3^0$. Calculate
\begin{eqnarray*}
\mbfM &=&  \mbfI_n + \nu_1 \sum_{l \in \calE_1} (\mbfe_{l_1}-\mbfe_{l_2})(\mbfe_{l_1}-\mbfe_{l_2})\trans \\
\mbfN &=& \nu_2 \sum_{k \in \calE_2} (\mbfe_{k_1}^*-\mbfe_{k_2}^*) (\mbfe_{k_1}^*-\mbfe_{k_2}^*)\trans + \nu_3 \bfone_p \bfone_p\trans
\end{eqnarray*}
For $m=1,2,\ldots$
\item Solve the Sylvester Equation $\mbfM\mbfA + \mbfA\mbfN = \mbfG^{m-1}$ to obtain $\mbfA^m$, where
\begin{eqnarray*}
\mbfG^{m-1}= \mbfX + \sum_{l \in \calE_1} (\mbfe_{l_1}-\mbfe_{l_2})(\blambda_{1l}^{m-1}+ \nu_1\mbfv_l^{m-1})\trans + \sum_{k \in \calE_2} (\blambda_{2k}^{m-1}+\nu_2\mbfz_k^{m-1})(\mbfe_{k_1}^*-\mbfe_{k_2}^*)\trans + \nu_3 \mbfs^{m-1} \bfone_p\trans,
\end{eqnarray*}
where $\mbfs^{m-1} = \bfone_n + \blambda_{3}^{m-1}/\nu_3$.
\item For $l\in\calE_1$, do
\begin{eqnarray*}
\mbfv_{l}^{m}= \textrm{prox}_{\sigma_{1l}\|\cdot\|_q} (A_{l_1\cdot}^{m}-A_{l_2\cdot}^{m}- \nu_1^{-1}\blambda_{1l}^{m-1}).
\end{eqnarray*}
\item For $k\in\calE_2$, do
\begin{eqnarray*}
\mbfz_{l}^{m}= \textrm{prox}_{\sigma_{2k}\|\cdot\|_q} (\mbfa_{k_1}^m-\mbfa_{k_2}^m - \nu_2^{-1}\blambda_{2k}^{m-1}).
\end{eqnarray*}
\item For $l\in\calE_1$ and $k \in\calE_2$, do
\begin{eqnarray*}
\blambda_{1l}^m &=& \blambda_{1l}^{m-1} + \nu_1(\mbfv_l^m - A_{l_1\cdot}^m + A_{l_2\cdot}^m)\\
\blambda_{2k}^m &=& \blambda_{2k}^{m-1} + \nu_2(\mbfz_k^m - \mbfa_{k_1}^m+\mbfa_{k_2}^m) \\
\blambda_{3}^m &=& \blambda_{3}^{m-1} + \nu_3(\bfone_n - \mbfA^m \bfone_p )
\end{eqnarray*}

\item Repeat Steps 2-5 until convergence. \end{enumerate}
\end{algorithm}

\section{Implementation of bi-ADMM and biC-ADMM in Practical Settings}

\label{sec:practical}

In this section, we discuss a few practical issues for implementing the proposed algorithms, including algorithmic convergence and selection of tuning parameters. The weights $w_l$ and $u_k$ are suggested in Section \ref{sec:genera_model}.

\subsection{Algorithmic Convergence}

The proposed convex biclustering algorithms are derived by applying standard techniques of the ADMM method. For convex clustering problems, \cite{Chi2015} and \cite{Wang2018sparse} as well as the references therein provided sufficient conditions for the convergence of their proposed methods. \cite{chi2017biconvex} discussed the convergence of their algorithm COBRA for convex biclustering problem.

The convergence of our bi-ADMM and biC-ADMM algorithms follows similar
arguments. Note that the only difference between the objective function
in \eqref{obj_general} and its counterpart in \citet{Chi2015}
is an additional fused-lasso penalty term for feature-level vectors, which is strongly convex. According
to \citet{Chi2015}, one can show that, under mild regularization conditions, the convergence
of the bi-ADMM and biC-ADMM algrithms is guaranteed for any $\nu_j>0, j=1,2,3$.

\subsection{Selection of Tuning Parameters}

\label{sec:tuning}

This subsection discusses the methods for selecting tuning parameters
$\gamma_{1}$ and $\gamma_{2}$. Recall that $\gamma_{1}$ controls
the number of estimated observation-level clusters and $\gamma_{2}$ controls the number of estimated feature-level clusters. The usage of two separate tuning parameters to control the numbers of row and column clusters separately provide great flexibility in applications.

Generally speaking, a two-dimensional grid search can provide an optimal pair of tuning parameters under some criterion if the computational resource is sufficient. Sometimes, if the data matrix is in a large scale, we may adopt a similar strategy suggested in \cite{chi2017biconvex} by using a single tuning parameter $\gamma$ for properly rescaled penalty terms. Using a single tuning parameter can reduce the computational burden, but row and columns have to be clustered in a proportional manner.

To be specific, we rewrite the convex biclustering problem as the following minimization problem with a single tuning parameter $\gamma$:
\begin{eqnarray} \label{obj_general_single}
\min_{\mbfA \in \mathbb{R}^{p \times n}} && \frac{1}{2} \sum_{i=1}^n\| X_{i\cdot}- A_{i\cdot} \|_2^2  +
\gamma \left\{\sum_{l \in \calE_1} w_l \| A_{l_1\cdot} -A_{l_2\cdot} \|_q +  \sum_{k \in \calE_2} u_k \| \mbfa_{k_1} -\mbfa_{k_2} \|_q \right\}.
\end{eqnarray}
The key to the validity of this formulation is that the two penalty terms $\sum_{l \in \calE_1} w_l \| A_{l_1\cdot} -A_{l_2\cdot} \|_q$ and $\sum_{k \in \calE_2} u_k \| \mbfa_{k_1} -\mbfa_{k_2} \|_q $ should be on the same scale. Note that row vectors of $\mbfA$ are in $\mathbb{R}^p$ while column vectors of $\mbfA$ are in $\mathbb{R}^n$. Thus, we choose row weights $w_l$ to sum to $1/\sqrt{p}$ and the column weights $u_k$ to sum to $1/\sqrt{n}$.

Next, we discuss two criteria for tuning parameter selection in a data-driven manner. One is spiritually similar with cross-validation and the other is based on stability selection. \cite{chi2017biconvex} proposed a hold-out validation method for convex biclustering by randomly selecting a hold-out of elements in the data matrix and assessing the quality of predicting the hold-out set with an estimated model based on the rest elements. Alternatively, \cite{Wang2018sparse} proposed to use stability selection in \citet{Fang2012} to tune two parameters for sparse convex clustering. For convex biclustering, we can apply stability section in a similar way to tune
$\gamma_{1}$ and $\gamma_{2}$. To be specific, for any given $\gamma_{1}$
and $\gamma_{2}$, two biclustering results can be produced via \eqref{obj_general} based on two sets of bootstrapped samples, and then the stability measurement \citep{Fang2012} can be computed to
measure the agreement between the two biclustering results. A repetition of this procedure for $50$ times is required to enhance the robustness of the stability selection method. Finally, the optimal parameters are selected as the one achieving maximum averaged stability value.

\section{Simulation Studies}

In this section, we conduct simulation studies to thoroughly evaluate the performance of the proposed bi-ADMM under general biclustering setups using $L_2$- and $L_1$-norms, and the biC-ADMM using $L_2$-norm under scenarios with compositional constrains motivated by a real microbiome dataset \cite{livanos2016antibiotic}, and compare them with the competing method COBRA. We do not illustrate the proposed algorithms with $L_\infty$-norm because extra computation burden is needed to solve the proximal map with simplex algorithms in Step 2 of Algorithms \ref{algbiADMM} and \ref{algbiCADMM}. For each setting, we run 100 repetitions. In all simulation studies, the adjusted RAND index (ARI) \citep{hubert1985comparing} is used to measure the agreement between the estimated clustering result and the underlying true clustering assignment. The ARI ranges from 0 to 1 with a higher value indicating better performance. Note that the underlying cluster labels are known or pre-specified in simulation studies, and thus, it is feasible to evaluate how well the candidate methods can perform if they are tuned by maximizing the ARI.

To ensure fair comparisons with the COBRA, we adopt the formulation shown in \eqref{obj_general_single}, using a single tuning parameter. For each repetition, an optimal tuning parameter $\gamma$ is chosen by maximizing the ARI on a separate validation data set for candidate algorithms.

\subsection{General Setups} \label{sec:sim_general}

To generate datasets with checkerboard cluster structures, we consider various sizes of datasets with $n$ observations (row) and $p$ features (column), where $(n,p)$ are set as $(50,40), (100, 80), (200,160)$, respectively. The number of row and column clusters varies as well. Given the dataset size, $K = 4,8,12$, or $16$ row clusters and $R = 4,8$, or $16$ column clusters are generated accordingly. The corresponding total number of biclusters are $M \doteq K \times R = 16, 32,64,96,128$, or $256$.

$X_{ij}$ is generated as follows. First, we assign cluster indexes to the observations (rows) by sampling from a set $\{1,..., K\}$ uniformly, and the cluster indexes are assigned to features (columns) following a similar procedure. Thus, each $X_{ij}$ belongs to one of those $M$ biclusters. Second, random samples for each bicluster are generated from a normal distribution,  $X_{ij} \text{ i.i.d. }\sim \calN(\mu_{kr}, \sigma^2)$, i.e., samples from the row cluster $k \in \{1,\ldots,K\}$ and column cluster $r\in \{1,\ldots,R\}$ follow a normal distribution with mean $\mu_{kr}$ and variance $\sigma^2$. $\mu_{kr}$ is chosen uniformly from a sequence $\{-10,-9,\ldots, 9, 10\}$. We vary the variance $\sigma^2$ to change the noise level, and the $\sigma$ is chosen as $2,4,6,8$, and $10$ in our simulation settings, respectively. Both parameters $\nu_1, \nu_2$ for the bi-ADMM algorithm are set as 8.

\begin{table}[!htb]
\LTcapwidth=\textwidth
\begin{longtable}[t]{cccccccccccc}
\caption{Simulation results for the bi-ADMM with $L_2$-norm, bi-ADMM with $L_1$-norm, and COBRA algorithms in terms of the ARI under various scenarios, respectively. The largest ARI(s) is bolded for each scenario.} \\
\toprule \toprule
\multicolumn{1}{c}{ } & \multicolumn{1}{c}{ } & \multicolumn{1}{c}{ } & \multicolumn{1}{c}{ } & \multicolumn{1}{c}{ } & \multicolumn{1}{c}{ } & \multicolumn{2}{c}{bi-ADMM ($L_2$)} & \multicolumn{2}{c}{bi-ADMM ($L_1$)} & \multicolumn{2}{c}{COBRA} \\
\cmidrule(l{2pt}r{2pt}){7-8} \cmidrule(l{2pt}r{2pt}){9-10} \cmidrule(l{2pt}r{2pt}){11-12}
$n$ & $p$ & $K$ & $R$ & $M$ & $\sigma$ & mean & sd & mean & sd & mean & sd \\
\midrule
\addlinespace[0.3em]
50 & 40 & 4 & 4 &16 & 2 & {\bf 1.00} & 0.03 & {\bf 1.00} & 0.02 & 0.99 & 0.05\\
         & & & & & 4 & {\bf 0.96} & 0.09 & {\bf 0.96} & 0.09 & 0.88 & 0.15\\
         & & & & & 6 & {\bf 0.85} & 0.16 & 0.83 & 0.18 & 0.66 & 0.19\\
         & & & & & 8 & {\bf 0.65} & 0.21 & 0.63 & 0.19 & 0.42 & 0.17\\
       & &8&4&32 & 2 & {\bf 0.99} & 0.02 & {\bf 0.99} & 0.03 & 0.97 & 0.06\\
         & & & & & 4 & {\bf 0.92} & 0.08 & {\bf 0.92} & 0.08 & 0.82 & 0.12\\
         & & & & & 6 & 0.76 & 0.12 & {\bf 0.77} & 0.11 & 0.57 & 0.15\\
         & & & & & 8 & 0.54 & 0.16 & {\bf 0.55} & 0.14 & 0.33 & 0.12\\
\addlinespace[0.3em]
100 & 80 &4&4& 16 & 2 & {\bf 1.00} & 0.02 & {\bf 1.00} & 0.03 & {\bf 1.00} & 0.02\\
          & & & & & 4 & {\bf 0.97} & 0.06 & 0.94 & 0.10 & 0.91 & 0.11\\
          & & & & & 6 & {\bf 0.86} & 0.15 & 0.75 & 0.19 & 0.84 & 0.20\\
          & & & & & 8 & {\bf 0.67} & 0.18 & 0.54 & 0.19 & 0.65 & 0.20\\
        & &8&4&32 & 4 & 0.91 & 0.09 & 0.91 & 0.09 & {\bf 0.93} & 0.08\\
          & & & & & 6 & 0.73 & 0.12 & 0.71 & 0.20 & {\bf 0.75} & 0.13\\
          & & & & & 8 & {\bf 0.53} & 0.12 & 0.51 & 0.14 & {\bf 0.53} & 0.14\\
        & &8&8&64 & 4 & {\bf 0.98} & 0.05 & 0.96 & 0.06 & {\bf 0.98} & 0.05\\
          & & & & & 6 & {\bf 0.82} & 0.17 & 0.78 & 0.13 & 0.81 & 0.13\\
          & & & & & 8 & 0.45 & 0.14 & {\bf 0.48} & 0.13 & 0.30 & 0.19\\
\addlinespace[0.3em]
200 & 160 &8&8& 64 & 4 & {\bf 1.00} & 0.02 & 0.99 & 0.03 & {\bf 1.00} & 0.02\\
           & & & & & 6 & {\bf 0.95} & 0.07 & 0.92 & 0.09 & {\bf 0.95} & 0.07\\
           & & & & & 8 & {\bf 0.80} & 0.13 & 0.71 & 0.15 & 0.77 & 0.14\\
           & & & & & 10 & {\bf 0.51} & 0.17 & 0.42 & 0.16 & 0.46 & 0.16\\
        & &12&8&256& 4 & {\bf 1.00} & 0.02 & {\bf 1.00} & 0.02 & {\bf 1.00} & 0.01\\
           & & & & & 6 & 0.95 & 0.05 & 0.94 & 0.06 & {\bf 0.97} & 0.05\\
           & & & & & 8 & {\bf 0.74} & 0.12 & 0.72 & 0.13 & 0.73 & 0.15\\
           & & & & & 10 & {\bf 0.46} & 0.14 & 0.40 & 0.13 & 0.39 & 0.13\\
        & &16&8&128& 4 & 0.99 & 0.03 & 0.99 & 0.02 & {\bf 1.00} & 0.01\\
           & & & & & 6 & {\bf 0.94} & 0.05 & 0.91 & 0.07 & 0.77 & 0.11\\
           & & & & & 8 & 0.66 & 0.11 & {\bf 0.69} & 0.11 & {\bf 0.69} & 0.12\\
           & & & & & 10 & {\bf 0.44} & 0.12 & 0.37 & 0.12 & 0.35 & 0.11\\
        & &16&16&256& 4 & {\bf 1.00} & 0.003 & {\bf 1.00} & 0.006 & {\bf 1.00} & 0.003\\
           & & & & & 6 & {\bf 0.98} & 0.03 & 0.95 & 0.05 & 0.96 & 0.04\\
           & & & & & 8 & {\bf 0.77} & 0.12 & 0.65 & 0.12 & 0.62 & 0.17\\
           & & & & & 10 & {\bf 0.40} & 0.15 & 0.23 & 0.08 & 0.28 & 0.10\\

\bottomrule \bottomrule \label{tab:1}
\end{longtable}
\end{table}

We compare the performance of bi-ADMM with $L_2$-norm (bi-ADMM ($L_2$)), bi-ADMM with $L_1$-norm (bi-ADMM ($L_1$)), and COBRA on identifying the bicluster structures in terms of the ARI. The mean value and the standard deviation of the ARI for each setting and algorithm across 100 replications are presented in Table \ref{tab:1}. The largest mean ARI value for each setting is bolded, and all tied values are bolded as well. Overall, we can see that as $\sigma$ increases, the ARIs for all three methods decrease. These observations are reasonable because a larger $\sigma$ implies a more challenging task to identify correct bicluster structures due to a higher noise level. In the meanwhile, given $\sigma$ and $M$, a larger sample size implies better performance. As to the differences between three methods, it is clear that bi-ADMM ($L_2$) outperform the others in most cases, while bi-ADMM ($L_1$) leads in a few cases with narrow margins. COBRA works comparably well in low-noise scenarios, but its performance deteriorates more quickly than bi-ADMM counterparts as the noise level increases, even worse than bi-ADMM ($L_1$) in some high-noise scenarios. The above results indicate that the proposed bi-ADMM methods are more robust convex biclustering algorithms. It is not surprising to see that bi-ADMM ($L_1$) is inferior to the other two methods in most cases, because the spherical clusters are generated under multivariate Gausiian distributions, which use $L_2$-norm internally.

\subsection{Compositional Data}

In this section, we evaluate the biclustering performance of biC-ADMM with $L_2$-norm and COBRA algorithms using simulated compositional data based on a murine gut microbiome study \citep{livanos2016antibiotic}, which compared the gut microbiome between the group with early-life pulsed therapeutic antibiotic dosing (PAT) treatment and the group without any antibiotic treatment (control). In accordance with prior microbiome data analysis \citep{koh2017powerful, hu2018two}, we first generate OTU counts for 100 subjects from the Dirichlet-multinomial (DM) distribution, where the dispersion parameter and proportion means were estimated from the control microbiome samples of the murine study, and the total number of reads per sample is set to 10,000. As a demonstration, a subset of the OTU count data consisting of 36 mice and 24 common taxa with average proportation $>0.01$ at age of 6 weeks is used in estimating the parameters of the DM distribution. Here the rule $>0.01$ helps to remove the extremely rare taxa and to facilitate the following manipulation. Before we manually manipulate one half of the generated samples to create biclusters, using the original data we sort the average proportions of taxa in an ascending order and assign them into 3 groups: Group 1 has the first six taxa with the smallest proportions, accounting for 3\% overall relative abundance on average; Group 2 has the last five taxa, accounting for 87\% overall relative abundance on average; Group 3 has the remaining 13 taxa, accounting for 10\% overall relative abundance on average. We manipulate the data in the following way. First, for each generated sample using the fitted DM model, we calculate the ratio of the summed counts for Group 2 over the summed counted for Group 1 (the averaged ratio is 126), and reduce the ratio by 1,400 folds. By doing that, we increased the relative abundance for taxa in Group 1 by roughly 93 folds from the original data. Second, we calculate the re-scaled counts for Groups 1 and 2 using the adjusted ratio by fixing the original total counts for Groups 1 and 2. Third, the count for each taxa in Groups 1 and 2 is re-scaled according to the ratio of the re-scaled group counts over the original ones. Lastly, the relative abundance for this sample is obtained by standardizing it using the total counts. By doing these, the relative abundance of the taxa in Group 3 remains unchanged, denoted as ``unchanged". Group 1 has its relative abundance enlarged for its taxa, while Group 2 has reduced relative abundance. Thus, Group 1 is denoted as ``enlarged", and Group 2 is denoted as ``shrunk". The 50 untouched samples are denoted as the ``control" group, and the 50 manipulated samples are denoted as the ``treatment" group. Then, the resulting data set has two row clusters and three column clusters.

Next, we evaluate the performance of COBRA and biC-ADMM algorithms on identifying the bicluster structure in terms of the ARI. The biculster structure with two row clusters and three column clusters may be artificial, so the ARI only serves as a reference metric. Under the compositional setting, a single tuning parameter $\gamma$ for the COBRA is tuned, using the default setting provided by the \textsf{R} package ``{\it cvxbiclustr}", and we adopt a two-dimensional grid search on $\gamma_1$ and $\gamma_2$ for the biC-ADMM method to allow for more flexibility.

\begin{figure}[!htb]
\protect\caption{Left panel: the heat map of the original simulated compositional data with 100 samples in 2 row clusters and 24 taxa in 3 column clusters. Right panel: the heat maps of $\wh\mbfA$ estimated by the biC-ADMM algorithm. The row(column) side color bars indicate the ``true" row(column) clusters.}
\centering \includegraphics[scale=0.5]{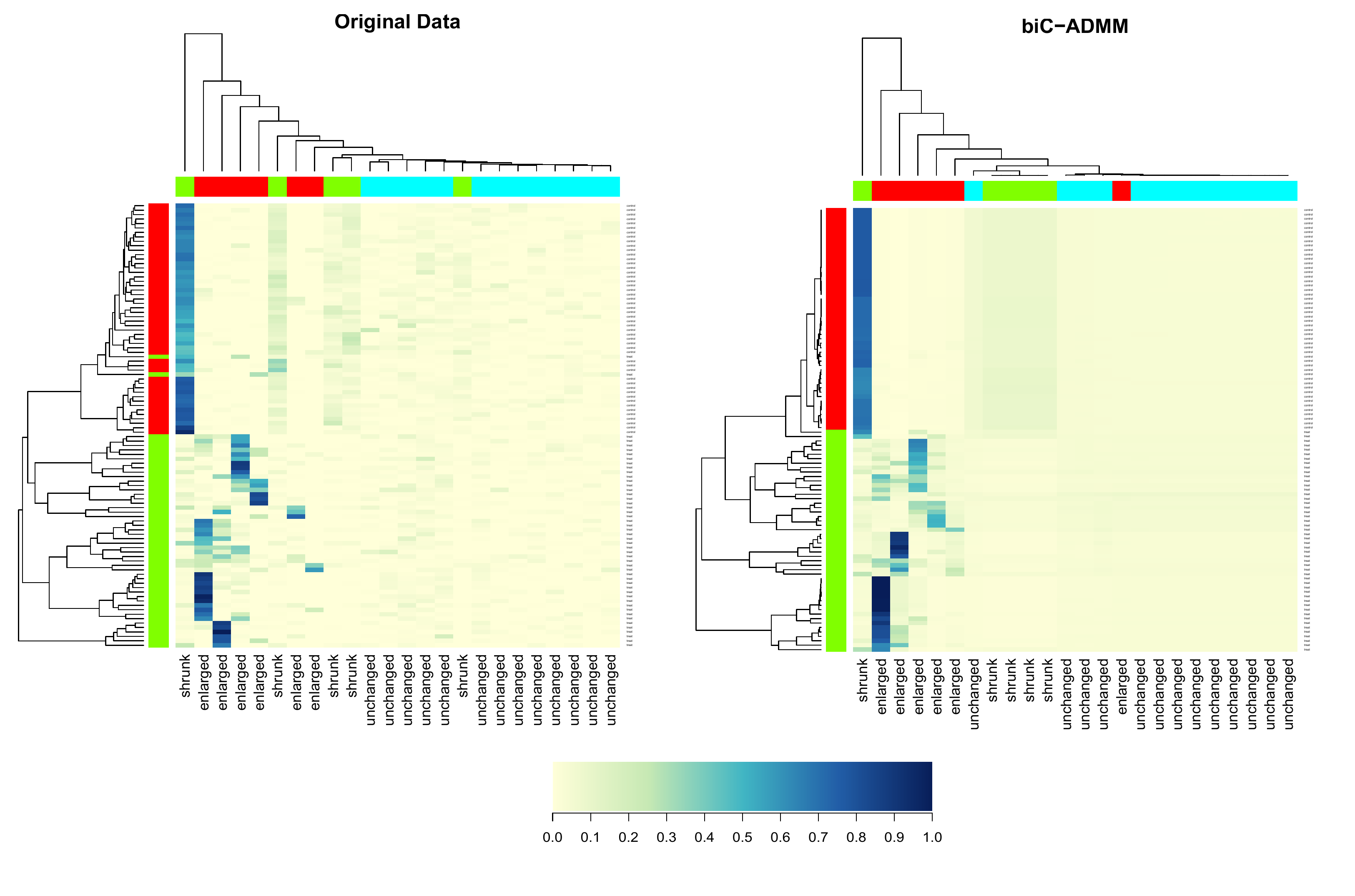}\label{sim_setting2_best}
\end{figure}

Table \ref{tab:2} shows that the biC-ADMM is superior to the COBRA algorithm in terms of the ARI. For one simulated data set, the best tuning parameters are $\gamma_1 = 37.28$ and $\gamma_2=268.27$. Figure \ref{sim_setting2_best} presents the corresponding heat map based on the estimated $\wh{\mbfA}$. We can see that the biC-ADMM algorithm correctly identifies the ``treatment" and ``control" groups for samples. For column clusters, the taxa from the same groups are almost classified in the same algorithm-generated clusters. The five taxa in the group ``shrunk" are included in two generated clusters with sizes as 1 and 4. The six taxa in the ``enlarged" group are split into two clusters also, with one mixed into a cluster primarily formed by taxa from the ``unchanged" group. Only one taxon from the ``unchanged" group is merged with taxa from the ``shrunk" group.

\begin{table}[!htb]
\begin{center}
\caption{Simulations results for biC-ADMM and COBRA algorithms in terms of the ARI under the compositional setup.}
\vspace{0.2cm}
\begin{tabular}{cccc}
\toprule \toprule
\multicolumn{2}{c}{biC-ADMM} & \multicolumn{2}{c}{COBRA} \\
\cmidrule(l{2pt}r{2pt}){1-2} \cmidrule(l{2pt}r{2pt}){3-4}
mean & sd & mean & sd\\
\midrule
\addlinespace[0.3em]
0.54 & 0.10 & 0.48 & 0.09 \\
\bottomrule \bottomrule \label{tab:2}
\end{tabular}
\end{center}
\end{table}

\subsection{Computational Consideration}

Although both the proposed Algorithm \ref{algbiADMM} and COBRA can target the same convex objective function \eqref{obj_general_single}, as we can see from Figure \ref{comparison}, the resulting estimates vary slightly. The dendrogram or an extra smoothing step can facilitate COBRA to convey the bicluster pattern, while our method can clearly display the pattern. One possible reason is that we directly derive the ADMM algorithm, and COBRA adopts an alternative approximation algorithm.

Step 2 in both Algorithms \ref{algbiADMM} and \ref{algbiCADMM} involves solving the {\it Sylvester Equation}, which imposes extra computational burden compared with the COBRA based on the DLPA and the AMA based convex clustering solver, which has less computational burden compared with the ADMM based solver \citep{Chi2015,Wang2018sparse}. As we implement the up-to-date algorithm to solve such type of equation, it is interesting to compare the computational time of the proposed methods to the COBRA on various data settings. In the following, we evaluate the computational time in seconds of the COBRA and the bi-ADMM algorithm under the $L_2$-norm. The COBRA is implemented with the $\textsf{R}$ package ``{\it cvxbiclustr}" \citep{chi2017biconvex}. We have developed an $\textsf{R}$ package ``{\it biADMM}" to implement our convex biclustering algorithms, bi-ADMM and biC-ADMM. We also developed a solver for the {\it Sylvester Equation}, which is based on a modified Bartels-Stewart algorithm proposed by \cite{Sorensen2003}. To improve the efficiency of the bi-ADMM algorithm, we also develop a Python version, which can be called back in $\textsf{R}$.

\begin{table}[!htb]
\centering
\LTcapwidth=\textwidth
\begin{longtable}[t]{lcccccc}
\caption{\label{tab:table-name} \fontsize{9}{11} Running time comparison with various combinations of $n, p, K, R$ under the setting described in Section \ref{sec:sim_general} with $\sigma = 2$ and a given tuning parameter $\gamma = 0.1$. The time is in seconds, which is an average across 30 replicates with 10 iterations for each algorithm.} \\
\toprule \toprule
& \multicolumn{1}{p{1.5cm}}{\centering $n = 50$ \\ $p = 40$ \\ $K = 4$ \\ $R = 4$}
& \multicolumn{1}{p{1.5cm}}{\centering $n = 50$ \\ $p = 40$ \\ $K = 8$ \\ $R = 4$}
& \multicolumn{1}{p{1.5cm}}{\centering $n = 100$ \\ $p = 80$ \\ $K = 4$ \\ $R = 4$}
& \multicolumn{1}{p{1.5cm}}{\centering $n = 100$ \\ $p = 80$ \\  $K = 8$ \\ $R = 4$}
& \multicolumn{1}{p{1.5cm}}{\centering $n = 200$ \\ $p = 160$ \\ $K = 12$ \\ $R = 8$}
& \multicolumn{1}{p{1.5cm}}{\centering $n = 200$ \\ $p = 160$ \\ $K = 16$ \\ $R = 8$} \\
\midrule
COBRA & 0.001 & 0.001 & 0.008 & 0.014 & 0.046 & 0.038\\
bi-ADMM &0.791 & 0.788 & 7.786 & 7.765 & 104.208 & 104.386\\
bi-ADMM (Python) &  0.072 & 0.075 & 0.657 & 0.617 & 6.247 & 6.336\\
\bottomrule \bottomrule  \label{tab:3}
\end{longtable}
\end{table}

The data sets are generated with the simulation settings described in Section \ref{sec:sim_general}. We set $\sigma = 2$ with a given tuning parameter $\gamma = 0.1$ defined in \eqref{obj_general_single}. Thirty replicates are simulated, and each algorithm is run for 10 iterations. Table \ref{tab:3} compares the average running time across 30 replicates of COBRA, bi-ADMM and Python-version bi-ADMM under various combinations of $n, p, K$ and $R$. The COBRA shows its superiority in program efficiency, with the shortest running time and the slowest increase as the scenario gets more complicated. The Python-version bi-ADMM program improves the running time significantly compared with the $\textsf{R}$ version. It is reasonable to pay acceptably more running time for better biclustering performance. The computer is equipped with a CPU i5-7267 (3.10 GHz) and 8G memory.

\section{Application of biC-ADMM to a Gut Microbiome Data}

\begin{figure}[!htb]
\caption{Left panel: the heat map of the original microbiome data with 68 samples and 37 taxa. Right panel: the estimate $\wh\mbfA$ based on the biC-ADMM algorithm. The row(column) side color bars indicate the row(column) clusters based on the algorithms.}
\centering \includegraphics[scale=0.45]{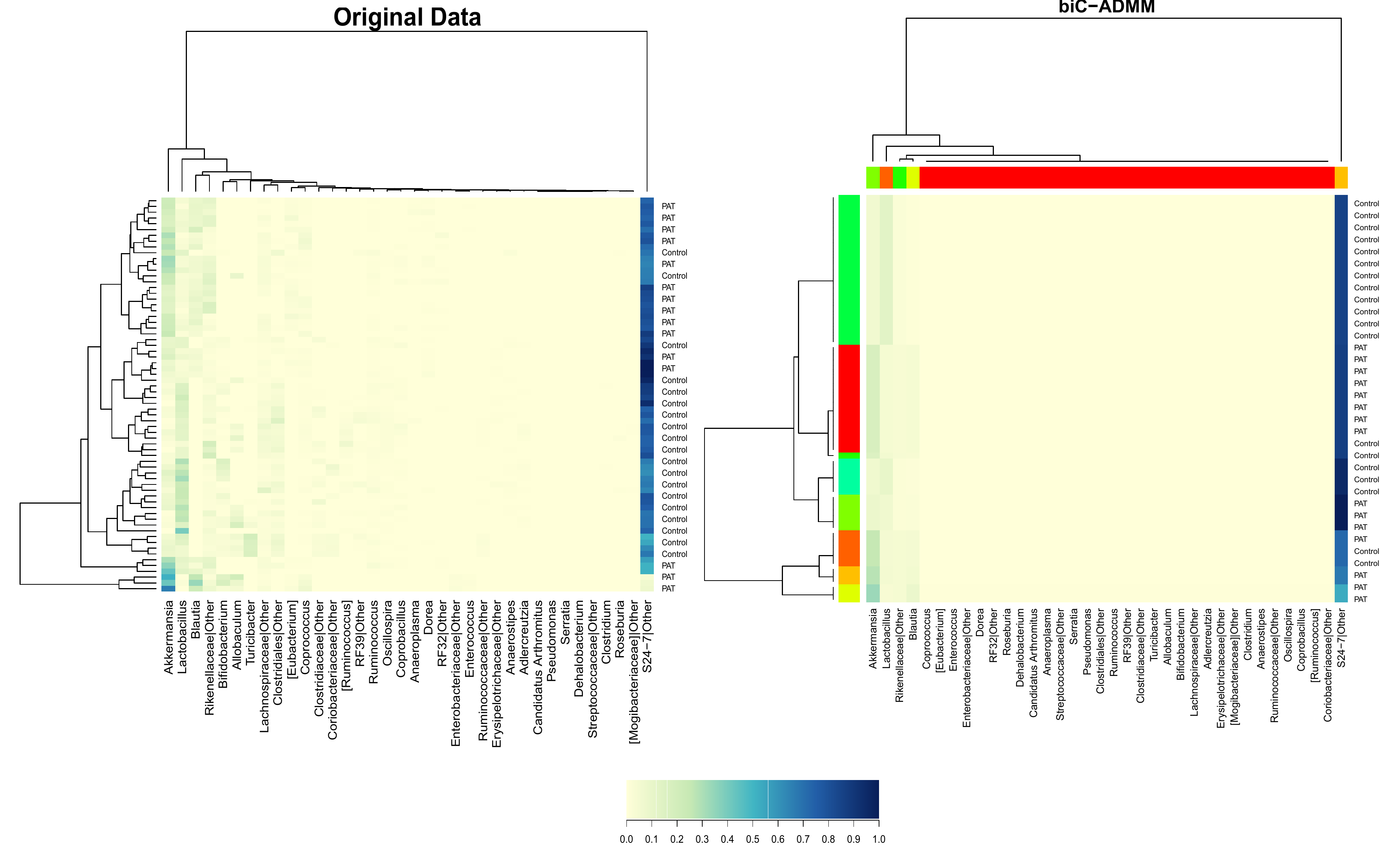} \label{fig:real_raw_best}
\end{figure}

Here, we apply the proposed biC-ADMM algorithm with two tuning parameters to the murine microbiome dataset used in simulation studies to assess the biclustering performance in compositional data. To be specific, we aim to simultaneously cluster a subset at 13 weeks with 37 common taxa and 68 samples, consisting of 32 microbiome samples from the PAT group and 36 control samples without retrieving their true treatment group information. The left panel of Figure \ref{fig:real_raw_best} shows the heat map and dendrogram of this data set. Since microbes in a community are usually dependent upon one another and may react similarly with environmental changes, it is of great interest to investigate whether there are multiple microbial taxa with the relative abundance being altered concordantly with the antibiotic treatment. In addition, the relative abundances of microbiome data are highly skewed, with a few common taxa accounting for the majority of bacteria in the sample. Microbiome studies usually need to group taxa at a higher rank to pull the signal strength. Some other research found that microbial community can be divided into sub-communities based on their functional reactions to the perturbation \citep{holmes2012dirichlet,sankaran2019latent}. All these observations motivate the application of clustering algorithm to identify data-driven taxa cluster patterns.

Following the commonly used pre-processing step in microbiome data analysis, the OTU count data of each sample is normalized into relative abundances prior to conducting biclustering, i.e., with relative abundances of 37 taxa summing up to 1. In our biC-ADMM algorithm, we set $m_1=m_2=5$ and $\phi=1$ for weight $w_{l}$ and $u_k$, and set $\nu_1=\nu_2=\nu_3=1$.

Then we apply the biC-ADMM algorithm to conduct convex biclustering for this data, which is an unsupervised learning method, free of a priori grouping knowledge. Although the true sample cluster labels are known, we mask them in biclustering analysis and aim to identify taxa clusters associated with potential sample subgroups. The tuning parameters are selected to maximize the stability for taxa clusters based on stability selection in Section \ref{sec:tuning}. The resulting optimal tuning parameters are $\gamma_1=153.99$ and $\gamma_2=82.54$. The bic-ADMM algorithm results into eight taxa clusters and seven sample clusters. It is noted that more sample clusters are obtained using biC-ADMM with an ARI of 0.40, but from Table \ref{tab:real_cross} we can see the purity of biC-ADMM generated clusters is high with only five samples are misclassified. The resulting taxa clusters indicate those taxa subgroups which have the potential to discriminate sample subgroups.

\begin{figure}[!htb]
\caption{Snap shots of the biC-ADMM solution path of the microbiome data, as $\gamma_1$ increases by row and $\gamma_2$ increase by column. }
\centering \includegraphics[scale=0.35]{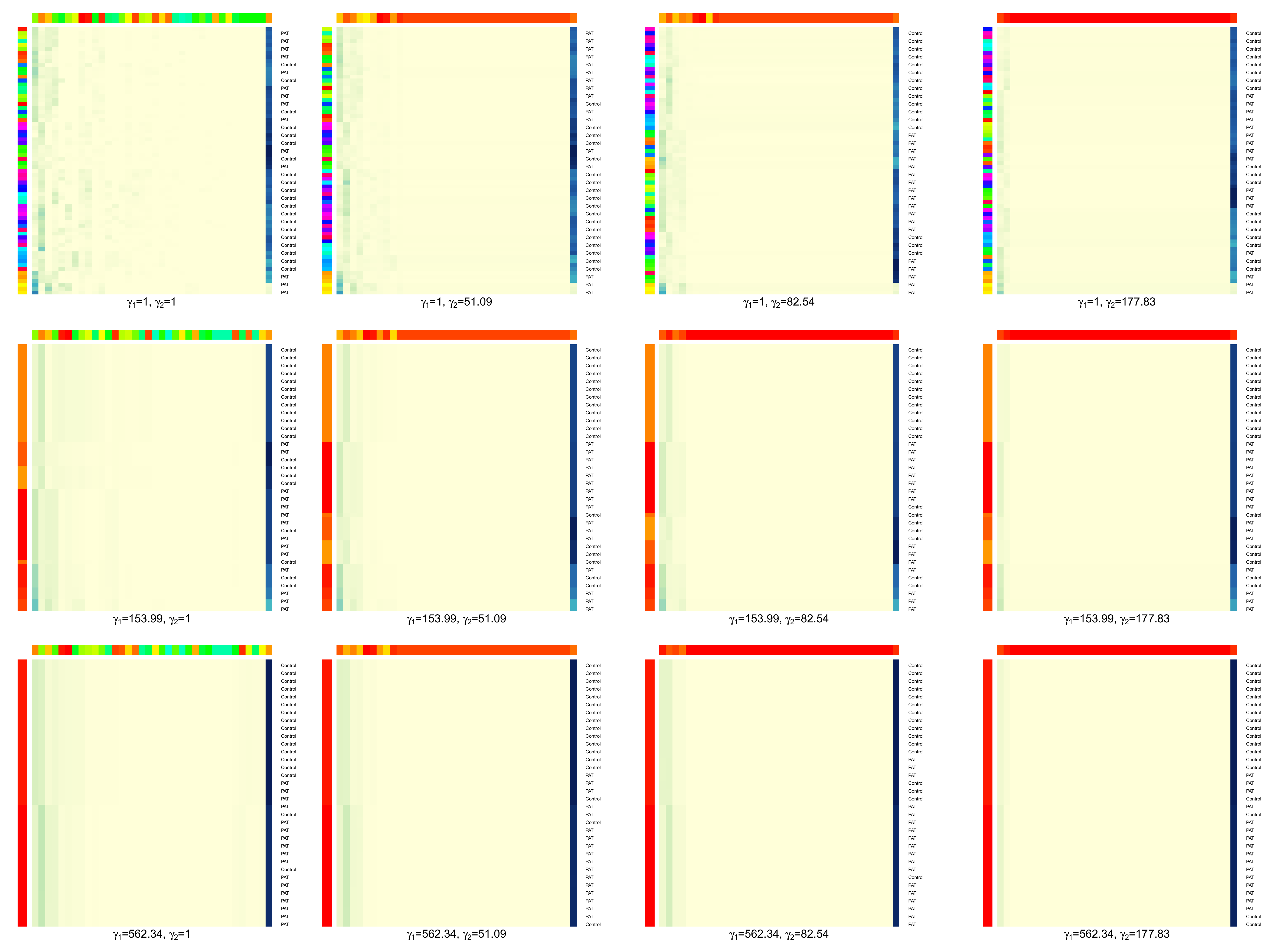} \label{fig:real_varying_gamma}
\end{figure}

Next, we illustrate how the solution $\wh\mbfA$ evolves as $\gamma_1$ and $\gamma_2$ vary. Figure \ref{fig:real_varying_gamma} shows snap shots of the biC-ADMM solution path of this data set, as the parameter $\gamma_1$ increases by row and $\gamma_2$ increases by column. The row path captures the whole range of behavior between under-smoothed estimates of the mean structure for taxa clusters (small $\gamma_2$) to over-smoothed estimates (large $\gamma_2$) with fixed sample clusters, while the row column captures the whole range of behavior between under-smoothed estimates for sample clusters (small $\gamma_1$) to over-smoothed estimates (large $\gamma_1$) with fixed taxa clusters. In between the extremes in top-left and bottom-right panels,
rows and columns are fusing together as $\gamma_1$ and $\gamma_2$ increases. The entire solution path enable us to identify those specific taxa that could discriminate sample groups. ``Akkermansia", ``Lactobacillus", ``Rikenellaceae|Other", ``Blautia" and ``S24-7|Other" seem to play an important role, and ``S24-7|Other" accounts for the largest relative abundance.

\begin{table}[!htb]
\begin{center}
\caption{The frequency table between true sample groups and clusters obtained from the biC-ADMM algorithm.}
\vspace{0.2cm}
\begin{tabular}{ccccccccc}
\toprule \toprule
& \multicolumn{8}{c}{biC-ADMM clusters} \\
\cmidrule(l{2pt}r{2pt}){2-9}
& 1 & 2 & 3 & 4 & 5 & 6 & 7 & 8\\
\midrule
\addlinespace[0.3em]
Control & 1 & 2 & 0 & 0 & 1 & 1 & 25 & 6  \\
PAT     & 17& 4 & 3 & 5 & 0 & 0 & 0  & 0 \\
\bottomrule \bottomrule \label{tab:real_cross}
\end{tabular}
\end{center}
\end{table}

\section{Summary} \label{sec:summary}

Biclustering is a powerful data mining technique which can provide results in a checkerboard-like pattern to explore the feature patterns for subgroups, while it is under-utilized in biological and biomedical data. Among biclustering algorithms developed in the past two decades, the convex biclustering algorithm is appealing due to its globally optimal solution. However, existing convex biclustering solved the convex optimization problem via an approximation to objective function of the ADMM algorithm, which required extra smoothing steps to obtain informative biclusters. On the other hand, existing biclustering methods cannot incorporate the compositional constraints which are often needed for the increasingly popular microbiome data.

In this manuscript, a new algorithm for the standard convex biclustering and its extension under the compositional constraints are proposed to simultaneously cluster observations and features, denoted as the bi-ADMM and the biC-ADMM, respectively. Both algorithms directly apply the ADMM method with solving a {\it Sylvester Equation} as a key step, which has a great potential to solve other problems with both row and columns constraints for a data matrix. The numerical results show that bi-ADMM and biC-ADMM are able to provide clear checkerboard-like pattern directly without any further smoothing step. Furthermore, the biC-ADMM can solve biclustering problems with compositional constraints, and it can be extended to any linear constraint easily, which has a great potential to be used in microbiome studies. Furthermore, we have developed an $\textsf{R}$ package ``{\it biADMM}" to implement our convex biclustering algorithms, bi-ADMM and biC-ADMM, which includes Python-version functions to speed up the calculation.

Moreover, this work can motivate future research. \cite{Wang2018sparse} presented the {\it Sparse Convex Clustering} problem, seeking to cluster observations and select features simultaneously. Feature selection can be incorporated into the convex biclustering as well. In addition, \cite{zhou2020efficient} developed a very efficient smoothing proximal gradient algorithm (Sproga) for convex clustering. Then, we can extend convex biclustering to sparse convex biclustering and develop a highly efficient algorithms to solve it, in order to conduct biclustering and feature selection simultaneously. Furthermore, weights $w_l$ and $u_k$ serves important roles in convex biclustering, but they suffer from the variable selection problem \citep{chakraborty2020biconvex}. The proposed algorithms can be improved in this direction as well.

\section*{Acknowledgement}

Hu and Li's research is partially supported by a National Institute of Health (NIH) grant (5R01DK110014).

\bibliographystyle{apalike}

\section*{Appendix}

\setcounter{equation}{0}
\global\long\def\theequation{A.\arabic{equation}}
 \global\long\def\thesubsection{A.\arabic{subsection}}
 \long\def\thelemma{A.\arabic{lemma}}
% \global\long\def\theproposition{A.\arabic{proposition}}

In Appendix, we provide proofs of Lemma \ref{thm:biADMM} and Lemma \ref{thm:biCADMM}.

\noindent {\bf Proof of Lemma \ref{thm:biADMM}}

Denote $\bfa=\textrm{vec}(\bfA)$, a vectorization of the matrix
$\bfA$. According to the fact that $A_{i_{1}\cdot}-A_{i_{2}\cdot}=\bfA\trans(\bfe_{i_{1}}-\bfe_{i_{2}})$, $\mbfa_{k_1}-\mbfa_{k_2}= \mbfA(\mbfe_{k_1}^*-\mbfe_{k2}^*)$,
and the property of the tensor product $\textrm{vec}(\bfR\bfS\bfT)=[\bfT\trans \otimes\bfR]\textrm{vec}(\bfS)$,
solving the minimization of $f(\bfA)$ is equivalent to minimize
\begin{eqnarray*}
f(\mbfa)=\frac{1}{2} \| \mbfx- \mbfa \|_2^2 + \frac{\nu_1}{2} \sum_{l \in \calE_l} \| \mbfB_l \mbfP \mbfa - \wt{\mbfv}_l  \|_2^2  +\frac{\nu_2}{2} \sum_{k \in \calE_2} \| \mbfC_k  \mbfa -\wt{\mbfz}_k \|_2^2 ,
\end{eqnarray*}
where $\mbfB_l, \mbfC_k$ and $\mbfP$ are defined in Section \ref{Alg_compositional}. Then it follows that
\begin{eqnarray*}
f(\mbfa)=\frac{1}{2} \| \mbfx- \mbfa \|_2^2 + \frac{\nu_1}{2} \| \mbfB \mbfP \mbfa - \wt{\mbfv}  \|_2^2  +\frac{\nu_2}{2}  \| \mbfC \mbfa -\wt{\mbfz} \|_2^2,
\end{eqnarray*}
where $\mbfB, \mbfC, \wt{\mbfv}$ and $\wt{\mbfz}$ are defined in Section \ref{Alg_compositional} as well.

Finally, the stationary equation can be obtained by
\begin{eqnarray*}
(\mbfI_{np} + \nu_1 \mbfP\trans \mbfB\trans \mbfB \mbfP + \nu_2 \mbfC\trans \mbfC ) \mbfa = \mbfx +\nu_1 \mbfP\trans \mbfB\trans \wt{\mbfv} + \nu_2 \mbfC\trans \wt{\mbfz}.
\end{eqnarray*}

This is a system of $np$ linear equations. By applying the properties for Kronecker product such as $(\mbfS\otimes \mbfT)\trans=\mbfS\trans \otimes \mbfT\trans$ and $(\mbfQ \otimes \mbfR)(\mbfS\otimes \mbfT)=(\mbfQ\mbfS) \otimes (\mbfR \mbfT)$, it follows
\begin{eqnarray*}
\nu_2 \mbfC\trans \mbfC &=& \nu_2 \sum_{k \in \calE_2} \left[ \left((\mbfe_{k_1}^*-\mbfe_{k_2}^*) (\mbfe_{k_1}^*-\mbfe_{k_2}^*)\trans \right) \right] \otimes \mbfI_n \\
&=& \left[\sum_{k \in \calE_2}\nu_2 \left((\mbfe_{k_1}^*-\mbfe_{k_2}^*) (\mbfe_{k_1}^*-\mbfe_{k_2}^*)\trans \right) \right] \otimes   \mbfI_n \\
\nu_2\mbfC\trans \wt{\mbfz} &=& \nu_2 \sum_{k \in \calE_2} [(\mbfe_{k_1}^*-\mbfe_{k_2}^*) \otimes \mbfI_n ] \wt{\mbfz}_k \\
&=& \sum_{k \in \calE_2}  \left(\nu_2(\mbfe_{k_1}^*-\mbfe_{k_2}^*)  \otimes  \mbfI_n \right) \wt{\mbfz}_k  \\
\nu_3 \mbfD\trans \mbfD &=& \nu_3 \sum_{i=1}^n \mbfD_i\trans \mbfD_i = \nu_3 (\bfone_p \bfone_p\trans) \otimes \mbfI_n \\
\nu_3\mbfD\trans\bfs &=& \nu_3 \sum_{i=1}^n (\bfone_p \otimes \mbfe_i) s_i = \nu_3 \sum_{i=1}^n s_i \textrm{vec}(\mbfe_i \bfone_p\trans ).
\end{eqnarray*}

In addition, by applying Proposition 1 in \cite{Wang2018sparse} for the permutation matrix $\mbfP$, we can prove
\begin{eqnarray*}
\mbfI_{np} + \nu_1 \mbfP\trans \mbfB\trans \mbfB \mbfP &=& \mbfI_p \otimes \left[ \mbfI_n + \nu_1 \sum_{l \in \calE_1} (\mbfe_{l_1}-\mbfe_{l_2})(\mbfe_{l_1}-\mbfe_{l_2})\trans \right] \\
\nu_1 \mbfP\trans \mbfB\trans \wt{\mbfv}&=&  \sum_{l \in \calE_1} \left(\mbfI_p \otimes \nu_1(\mbfe_{l_1}-\mbfe_{l_2}) \right) \wt{\mbfv}_l .
\end{eqnarray*}

Therefore, the system of equations is equivalent to
\begin{eqnarray*}
(\mbfI_p \otimes \mbfM)\textrm{vec}(\mbfA) + (\mbfN\otimes \mbfI_n )\textrm{vec}(\mbfA) = \textrm{vec}(\mbfG),
\end{eqnarray*}
which is equivalent to $\mbfM\mbfA + \mbfA\mbfN = \mbfG$. \hfill $\Box$

\bigskip

\noindent {\bf Proof of Lemma \ref{thm:biCADMM}}

Proof of Lemma \ref{thm:biADMM} needs to be modified to incorporate the extra constraints.
The stationary equations for minimizing \eqref{eqn:biC_update_A} can be obtained by
\begin{eqnarray*}
(\mbfI_{np} + \nu_1 \mbfP\trans \mbfB\trans \mbfB \mbfP + \nu_2 \mbfC\trans \mbfC + \nu_3 \mbfD\trans \mbfD) \mbfa = \mbfx +\nu_1 \mbfP\trans \mbfB\trans \wt{\mbfv} + \nu_2 \mbfC\trans \wt{\mbfz} + \nu_3 \mbfD\trans\bfs.
\end{eqnarray*}

With matrix techniques, it follows
\begin{eqnarray*}
\nu_3 \mbfD\trans \mbfD &=& \nu_3 \sum_{i=1}^n \mbfD_i\trans \mbfD_i = \nu_3 (\bfone_p \bfone_p\trans) \otimes \mbfI_n \\
\nu_3\mbfD\trans\bfs &=& \nu_3 \sum_{i=1}^n (\bfone_p \otimes \mbfe_i) s_i = \nu_3 \sum_{i=1}^n s_i \textrm{vec}(\mbfe_i \bfone_p\trans ).
\end{eqnarray*}

Therefore, the system of equations is equivalent to
$$(\mbfI_p \otimes \mbfM)\textrm{vec}(\mbfA) + (\mbfN\otimes \mbfI_n )\textrm{vec}(\mbfA) = \textrm{vec}(\mbfG).$$
\hfill $\Box$

\end{document}

%% file: self-define-YF.tex
\def\wt{\widetilde}

\def\wh{\widehat}

\def\boxit#1{\vbox{\hrule\hbox{\vrule\kern6pt
          \vbox{\kern6pt#1\kern6pt}\kern6pt\vrule}\hrule}}

\def\bse{\begin{eqnarray*}}
\def\ese{\end{eqnarray*}}
\def\be{\begin{eqnarray}}
\def\ee{\end{eqnarray}}
\def\bq{\begin{equation}}
\def\eq{\end{equation}}

\def\wh{\widehat}

\def\trans{^{\rm T}}

\newcommand{\blem}{\begin{lemma}}
\newcommand{\elem}{\end{lemma}}
\newcommand{\bthe}{\begin{theorem}}
\newcommand{\ethe}{\end{theorem}}

\newtheorem{lemma}{Lemma}
\newtheorem{theorem}{Theorem}
\def\bfa{{\bf a}}
\def\bfA{{\bf A}}

\def\bfe{{\bf e}}

\def\bfI{{\bf I}}

\def\bfR{{\bf R}}
\def\bfs{{\bf s}}
\def\bfS{{\bf S}}

\def\bfT{{\bf T}}
\def\bfu{{\bf u}}

\def\bfv{{\bf v}}

\def\bfX{{\bf X}}

\def\blambda{{\mbox{\boldmath $\lambda$}}}
\def\bLambda{{\mbox{\boldmath $\Lambda$}}}

\def\delete#1{\iffalse #1 \fi}

\def\bse{\begin{eqnarray*}}
\def\ese{\end{eqnarray*}}
\def\bee{\begin{enumerate}}
\def\eee{\end{enumerate}}
\def\bqe{\begin{eqnarray}}
\def\eqe{\end{eqnarray}}
\def\bed{\begin{description}}
\def\eed{\end{description}}
\def\bei{\begin{itemize}}
\def\eei{\end{itemize}}

\def\argmin{\mathop{\rm argmin}}

\def\pmb#1{\setbox0=\hbox{#1}%
    \kern-.025em\copy0\kern-\wd0
    \kern.05em\copy0\kern-\wd0
    \kern-.025em\raise.0433em\box0 }
\def\pmbh#1#2{\setbox0=\hbox{#1}%
    \setbox1=\hbox{#2}%
    \kern-.025em\copy0\kern-\wd0
    \kern.05em\copy1\kern-\wd0
    \kern-.025em\raise.0433em\box0 }

\def\frac#1#2{{#1\over#2}}

\def\boxit#1{\vbox{\hrule\hbox{\vrule\kern6pt
   \vbox{\kern6pt#1\kern6pt}\kern6pt\vrule}\hrule}}

\def\listing#1{\vskip 4mm\begin{verbatim}\input#1 \vskip 4mm}
\def\thick#1{\hbox{\rlap{$#1$}\kern0.25pt\rlap{$#1$}\kern0.25pt$#1$}}

\def\wt{\widetilde}
\def\wh{\widehat}

\def\bfa{{\bf a}}  \def\bfA{{\bf A}}

\def\bfe{{\bf e}}

  \def\bfI{{\bf I}}

  \def\bfR{{\bf R}}
\def\bfs{{\bf s}}  \def\bfS{{\bf S}}
  \def\bfT{{\bf T}}
\def\bfu{{\bf u}}  
\def\bfv{{\bf v}}  
  
  \def\bfX{{\bf X}}

\def\bfzero{{\bf 0}}
\def\bfone{{\bf 1}}
\def\pmbh{{\pmb h}}

\def\calE{{\cal E}}

\def\calL{{\cal L}}

\def\calN{{\cal N}}

\def\mbfa{\mathbf{a}}  \def\mbfA{\mathbf{A}}
  \def\mbfB{\mathbf{B}}
  \def\mbfC{\mathbf{C}}
  \def\mbfD{\mathbf{D}}
\def\mbfe{\mathbf{e}}  
  
  \def\mbfG{\mathbf{G}}
  
  \def\mbfI{\mathbf{I}}

  \def\mbfM{\mathbf{M}}
  \def\mbfN{\mathbf{N}}

  \def\mbfP{\mathbf{P}}
  \def\mbfQ{\mathbf{Q}}
  \def\mbfR{\mathbf{R}}
\def\mbfs{\mathbf{s}}  \def\mbfS{\mathbf{S}}
  \def\mbfT{\mathbf{T}}
  
\def\mbfv{\mathbf{v}}  \def\mbfV{\mathbf{V}}
  
\def\mbfx{\mathbf{x}}  \def\mbfX{\mathbf{X}}
  
\def\mbfz{\mathbf{z}}  \def\mbfZ{\mathbf{Z}}

\renewcommand\today{\ifcase\month\or
   Jan\or Feb\or Mar\or Apr\or May\or
   Jun\or Jul\or Aug\or Sep\or Oct\or Nov\or
   Dec\fi
   \space\number\day, \number\year}

%% file: biclustering_20210608.bbl
\begin{thebibliography}{}

\bibitem[Bartels and Stewart, 1972]{bartels1972solution}
Bartels, R.~H. and Stewart, G.~W. (1972).
\newblock Solution of the matrix equation ax+ xb= c [f4].
\newblock {\em Communications of the ACM}, 15(9):820--826.

\bibitem[Busygin et~al., 2008]{busygin2008biclustering}
Busygin, S., Prokopyev, O., and Pardalos, P.~M. (2008).
\newblock Biclustering in data mining.
\newblock {\em Computers \& Operations Research}, 35(9):2964--2987.

\bibitem[Chakraborty and Xu, 2020]{chakraborty2020biconvex}
Chakraborty, S. and Xu, J. (2020).
\newblock Biconvex clustering.

\bibitem[Cheng and Church, 2000]{cheng2000biclustering}
Cheng, Y. and Church, G.~M. (2000).
\newblock Biclustering of expression data.
\newblock In {\em Ismb}, volume~8, pages 93--103.

\bibitem[Chi et~al., 2017]{chi2017biconvex}
Chi, E.~C., Allen, G.~I., and Baraniuk, R.~G. (2017).
\newblock Convex biclustering.
\newblock {\em Biometrics}, 73(1):10--19.

\bibitem[Chi and Lange, 2015]{Chi2015}
Chi, E.~C. and Lange, K. (2015).
\newblock {Splitting Methods for Convex Clustering}.
\newblock {\em Journal of Computational and Graphical Statistics},
  24(4):994--1013.

\bibitem[Falony et~al., 2016]{falony2016population}
Falony, G., Joossens, M., Vieira-Silva, S., Wang, J., Darzi, Y., Faust, K.,
  Kurilshikov, A., Bonder, M.~J., Valles-Colomer, M., Vandeputte, D., et~al.
  (2016).
\newblock Population-level analysis of gut microbiome variation.
\newblock {\em Science}, 352(6285):560--564.

\bibitem[Fang and Wang, 2012]{Fang2012}
Fang, Y. and Wang, J. (2012).
\newblock {Selection of the number of clusters via the bootstrap method}.
\newblock {\em Computational Statistics {\&} Data Analysis}, 56:468--477.

\bibitem[Flynn and Perry, 2012]{flynn2012consistent}
Flynn, C.~J. and Perry, P.~O. (2012).
\newblock Consistent biclustering.
\newblock {\em arXiv preprint arXiv:1206.6927}, 3.

\bibitem[Friedman et~al., 2001]{hastie2009}
Friedman, J., Hastie, T., and Tibshirani, R. (2001).
\newblock {\em The elements of statistical learning}, volume~1.
\newblock Springer series in statistics New York.

\bibitem[Gloor et~al., 2017]{gloor2017microbiome}
Gloor, G.~B., Macklaim, J.~M., Pawlowsky-Glahn, V., and Egozcue, J.~J. (2017).
\newblock Microbiome datasets are compositional: and this is not optional.
\newblock {\em Frontiers in microbiology}, 8:2224.

\bibitem[Hochreiter et~al., 2010]{hochreiter2010fabia}
Hochreiter, S., Bodenhofer, U., Heusel, M., Mayr, A., Mitterecker, A., Kasim,
  A., Khamiakova, T., Van~Sanden, S., Lin, D., Talloen, W., et~al. (2010).
\newblock Fabia: factor analysis for bicluster acquisition.
\newblock {\em Bioinformatics}, 26(12):1520--1527.

\bibitem[Holmes et~al., 2012]{holmes2012dirichlet}
Holmes, I., Harris, K., and Quince, C. (2012).
\newblock Dirichlet multinomial mixtures: generative models for microbial
  metagenomics.
\newblock {\em PloS one}, 7(2):e30126.

\bibitem[Hripcsak and Albers, 2012]{hripcsak2012next}
Hripcsak, G. and Albers, D.~J. (2012).
\newblock Next-generation phenotyping of electronic health records.
\newblock {\em Journal of the American Medical Informatics Association},
  20(1):117--121.

\bibitem[Hu et~al., 2018]{hu2018two}
Hu, J., Koh, H., He, L., Liu, M., Blaser, M.~J., and Li, H. (2018).
\newblock A two-stage microbial association mapping framework with advanced fdr
  control.
\newblock {\em Microbiome}, 6(1):131.

\bibitem[Hubert and Arabie, 1985]{hubert1985comparing}
Hubert, L. and Arabie, P. (1985).
\newblock Comparing partitions.
\newblock {\em Journal of classification}, 2(1):193--218.

\bibitem[Jameson, 1968]{Jameson1968}
Jameson, A. (1968).
\newblock Solution of the equation ax+xb=c by inversion of an m*m or n*n
  matrix.
\newblock {\em SIAM Journal on Applied Mathematics}, 16(5):1020--1023.

\bibitem[Koh et~al., 2017]{koh2017powerful}
Koh, H., Blaser, M.~J., and Li, H. (2017).
\newblock A powerful microbiome-based association test and a microbial taxa
  discovery framework for comprehensive association mapping.
\newblock {\em Microbiome}, 5(1):45.

\bibitem[Lee et~al., 2010]{Lee2010}
Lee, M., Shen, H., Huang, J.~Z., and Marron, J.~S. (2010).
\newblock {Biclustering via Sparse Singular Value Decomposition}.
\newblock {\em Biometrics}, 66(4):1087--1095.

\bibitem[Li, 2020]{li2020generalized}
Li, G. (2020).
\newblock Generalized co-clustering analysis via regularized alternating least
  squares.
\newblock {\em Computational statistics \& data analysis}, 150:106989.

\bibitem[Livanos et~al., 2016]{livanos2016antibiotic}
Livanos, A.~E., Greiner, T.~U., Vangay, P., Pathmasiri, W., Stewart, D.,
  McRitchie, S., Li, H., Chung, J., Sohn, J., Kim, S., et~al. (2016).
\newblock Antibiotic-mediated gut microbiome perturbation accelerates
  development of type 1 diabetes in mice.
\newblock {\em Nature microbiology}, 1(11):1--13.

\bibitem[Madeira and Oliveira, 2004]{madeira2004biclustering}
Madeira, S.~C. and Oliveira, A.~L. (2004).
\newblock Biclustering algorithms for biological data analysis: a survey.
\newblock {\em IEEE/ACM Transactions on Computational Biology and
  Bioinformatics (TCBB)}, 1(1):24--45.

\bibitem[Sankaran and Holmes, 2019]{sankaran2019latent}
Sankaran, K. and Holmes, S.~P. (2019).
\newblock Latent variable modeling for the microbiome.
\newblock {\em Biostatistics}, 20(4):599--614.

\bibitem[Shabalin et~al., 2009]{shabalin2009finding}
Shabalin, A.~A., Weigman, V.~J., Perou, C.~M., and Nobel, A.~B. (2009).
\newblock Finding large average submatrices in high dimensional data.
\newblock {\em The Annals of Applied Statistics}, pages 985--1012.

\bibitem[Sorensen et~al., 2003]{Sorensen2003}
Sorensen, D.~C., Zhou, Y., et~al. (2003).
\newblock Direct methods for matrix sylvester and lyapunov equations.
\newblock {\em Journal of Applied Mathematics}, 6(2003):277--303.

\bibitem[Tan and Witten, 2014]{Tan2014}
Tan, K.~M. and Witten, D.~M. (2014).
\newblock Sparse biclustering of transposable data.
\newblock {\em Journal of Computational and Graphical Statistics},
  23(4):985--1008.
\newblock PMID: 25364221.

\bibitem[Tibshirani et~al., 2005]{Tibshirani.ea:2005}
Tibshirani, R., Saunders, M., Rosset, S., Zhu, J., and Knight, K. (2005).
\newblock Sparsity and smoothness via the fused lasso.
\newblock {\em Journal of the Royal Statistical Society: Series B},
  67:1198--1232.

\bibitem[Wang et~al., 2018]{Wang2018sparse}
Wang, B., Zhang, Y., Sun, W.~W., and Fang, Y. (2018).
\newblock Sparse convex clustering.
\newblock {\em Journal of Computational and Graphical Statistics}, 27:393--403.

\bibitem[Xie et~al., 2018]{xie2018time}
Xie, J., Ma, A., Fennell, A., Ma, Q., and Zhao, J. (2018).
\newblock It is time to apply biclustering: a comprehensive review of
  biclustering applications in biological and biomedical data.
\newblock {\em Briefings in bioinformatics}, 1:16.

\bibitem[Zhou et~al., 2020]{zhou2020efficient}
Zhou, X., Du, C., and Cai, X. (2020).
\newblock An efficient smoothing proximal gradient algorithm for convex
  clustering.
\newblock {\em arXiv preprint arXiv:2006.12592}.

\end{thebibliography}
